\documentclass[final,3p,times]{elsarticle}

\usepackage{amssymb}
\usepackage{amsmath}
\usepackage{upgreek}
\usepackage{float}
\usepackage{lineno} 
\usepackage{colortbl} 
\definecolor{grayshade}{rgb}{0.7,0.7,0.7} 

\journal{TBD}

\begin{document}

\begin{frontmatter}

\title{Two-Shot Optimization of Compositionally Complex Refractory Alloys}

\author[aff1,aff2]{James D. Paramore\corref{cor1}}
\ead{james.paramore@tamu.edu}
\author[aff3]{Brady G. Butler}
\author[aff1]{Michael T. Hurst}
\author[aff2]{Trevor Hastings}
\author[aff2]{Daniel O. Lewis}
\author[aff2]{Eli Norris}
\author[aff2]{Benjamin Barkai}
\author[aff2]{Joshua Cline}
\author[aff2]{Braden Miller}
\author[aff2]{Jose Cortes}
\author[aff2]{Ibrahim Karaman}
\author[aff2]{George M. Pharr}
\author[aff2]{Raymundo Arroyave\corref{cor1}}
\ead{raymundo.arroyave@tamu.edu}
\cortext[cor1]{Corresponding authors}

\affiliation[aff1]
            {
            organization={Bush Combat Development Complex, Texas A\&M University System},
            addressline={3479 TAMU}, 
            city={College Station},
            state={TX 77843-3479},
            country={USA}
            }

\affiliation[aff2]
            {
            organization={Department of Materials Science and Engineering, Texas A\&M University},
            addressline={3003 TAMU}, 
            city={College Station},
            state={TX 77843-3003},
            country={USA}
            }

\affiliation[aff3]
            {
            organization={DEVCOM Army Research Laboratory},
            addressline={3003 TAMU}, 
            city={College Station},
            state={TX 77843-3003},
            country={USA}
            }

\begin{abstract}
In this paper, a synergistic computational/experimental approach is presented for the rapid discovery and characterization of novel alloys within the compositionally complex (i.e., “medium/high entropy”) refractory alloy space of Ti-V-Nb-Mo-Hf-Ta-W. This was demonstrated via a material design cycle aimed at simultaneously maximizing the objective properties of high specific hardness (hardness normalized by density) and high specific elastic modulus (elastic modulus normalized by density). This framework utilizes high-throughput computational thermodynamics and intelligent filtering to first reduce the untenably large alloy space to a feasible size, followed by an iterative design cycle comprised of high-throughput synthesis, processing, and characterization in batch sizes of 24 alloys. After the first iteration, Bayesian optimization was utilized to inform selection of the next batch of 24 alloys. This paper demonstrates the benefit of using batch Bayesian optimization (BBO) in material design, as significant gains in the objective properties were observed after only two iterations or “shots” of the design cycle without using any prior knowledge or physical models of how the objective properties relate to the design inputs (i.e., composition). Specifically, the hypervolume of the Pareto front increased by 54\% between the first and second iterations. Furthermore, 10 of the 24 alloys in the second iteration dominated all alloys from the first iteration.
\end{abstract}




\end{frontmatter}


\section{Introduction}
\label{Introduction}
\begin{figure}
    \centering
    \includegraphics[width=0.4\linewidth]{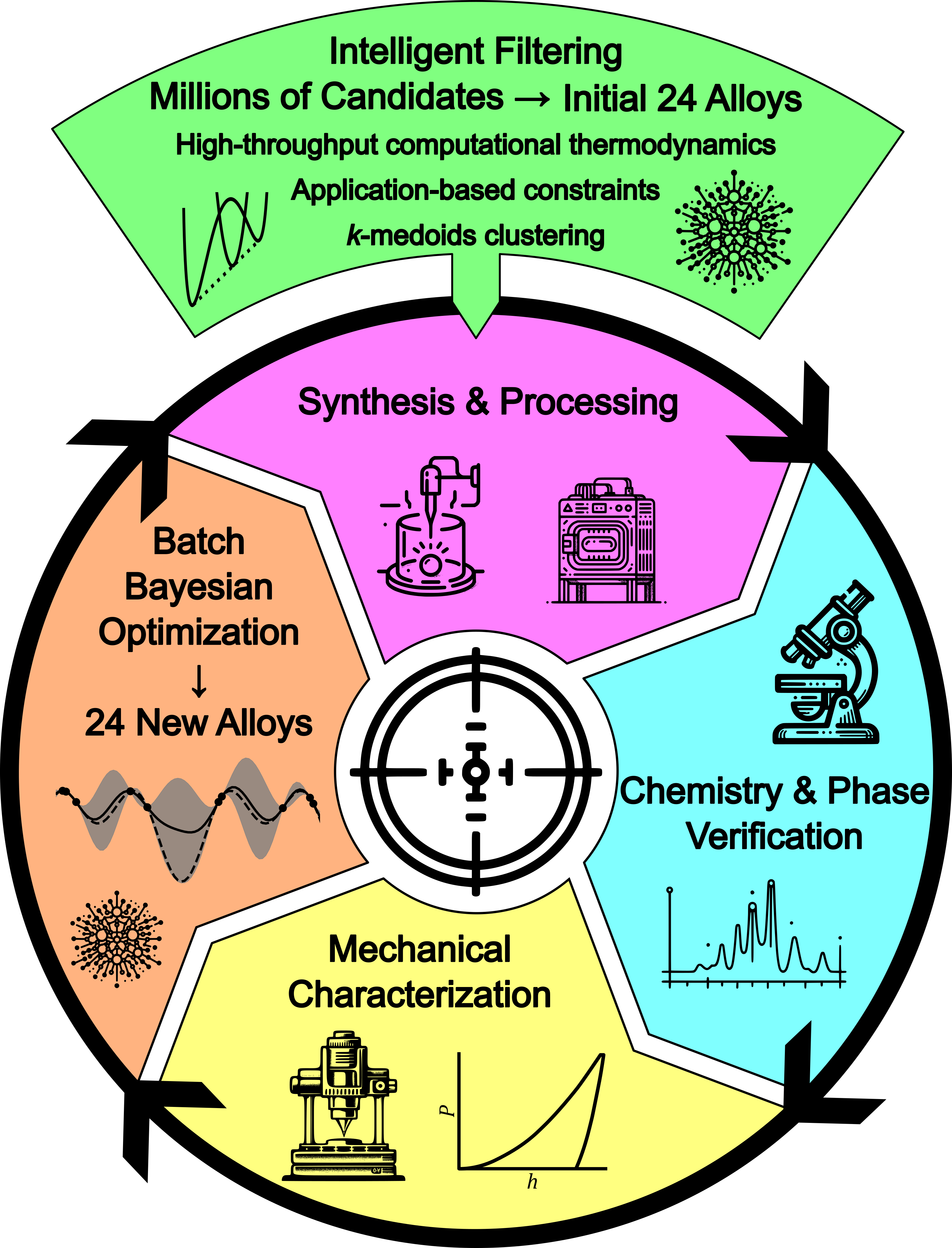}
    \caption{Computational and experimental workflow for the iterative design process used in this work.}
    \label{fig:Workflow}
\end{figure}

There is a considerable delay from the initial conception of new technologies to the development of materials with the necessary properties to realize them. To meet society’s most pressing modernization priorities, traditional materials discovery is far too slow and, therefore, insufficient. Specific challenges include: (1) how to quickly assess the feasibility of a region in a chemical space under multiple performance constraints; (2) how to synthesize candidate materials in a high‐throughput manner; (3) how to efficiently assess the performance of these materials at a sufficiently fast rate to support effective decision making; and (4) how to integrate the design, synthesis, and characterization steps in an iterative loop that facilitates an optimal balance of exploration for new data and exploitation of existing data within promising, but extremely large, materials design spaces. To address the challenges above, this work implemented a framework for closed‐loop design, synthesis, and characterization of materials. The framework developed and used in this work is based on the BIRDSHOT (Batch-wise Improvement in Reduced Design Space using a Holistic Optimization Technique) approach \cite{Khatamsaz2023,Couperthwaite2020,Vela2023,Vela2023-2}. A schematic depiction of the iterative workflow used in this study is given Figure \ref{fig:Workflow}.

While no specific application was envisioned for this work, a generic application was assumed in which the components are subjected to both extremely high temperatures and violent mechanical loading events (e.g., rapid accelerations, vibrations, shock). To this end, a refractory metal alloy system (Ti-V-Nb-Mo-Hf-Ta-W) was chosen as the candidate space. As discussed in section \ref{Initial Filtering}, candidate alloys were filtered to only those that had ideal configurational entropy greater than or equal to equimolar ternary systems (i.e., $S_\mathrm{conf}^\mathrm{ideal} \ge R \ln 3$, where $R$ is the gas constant). This constraint puts all feasible alloys within the “high entropy alloy” (HEA) or “medium entropy alloy” (MEA) ranges by one or more previously prescribed definitions \cite{Miracle2017,Yeh2006,George2019}. This is mentioned to help the reader understand that the current work is related to this popular field of study in alloy design. The authors understand that quantifying or labeling the true entropy of an alloy requires more consideration than overall composition. However, such considerations are well beyond the scope of this work. Therefore, “compositionally complex refractory alloys” is the chosen terminology in this manuscript. Regardless of semantics, such materials are attractive for extreme environment applications due to the potential for phases with high configurational entropy \cite{George2019,Senkov2018}. This, in turn, can lead to high thermodynamic stability against various degradation mechanisms, including high temperature/pressure, mechanical vibration/shock, and corrosion. Furthermore, the compositional complexity results in vast candidate spaces with corresponding freedom and flexibility in exploration, allowing for simultaneous optimization of multiple objective properties.

The objective properties for optimization in this work were chosen as high specific hardness (hardness normalized by density) and high specific elastic modulus (elastic modulus normalized by density), as measured by nanoindentation. Specific hardness, as an indication of specific strength, and specific modulus were chosen because they are (1) well suited to high-throughput measurements via nanoindentation using the authors’ pre-existing capabilities and expertise and (2) the mechanical performance of a material subjected to extreme accelerations, vibrations, and/or shock is directly related to these properties. For a given acceleration, whether linear, non-linear, or oscillatory, denser materials will be subjected to greater stresses. Furthermore, any elastic or plastic deformation of that material due to stresses experienced during such events will be related to its modulus and strength (or hardness), respectively. Therefore, optimizing hardness and modulus normalized by density should identify materials promising for their resistance to deformation or failure due to rapid accelerations.

While a specific material system (i.e., Ti-V-Nb-Mo-Hf-Ta-W alloys) and specific mechanical properties (i.e., indentation response) were chosen for optimization in this work, it should be emphasized that the computational aspects of this framework can be used for \textit{\textbf{any}} material system in \textbf{\textit{any}} material class (metals, ceramics, polymers, composites, etc.) and can be used to optimize \textbf{\textit{any}} quantifiable material property (mechanical, thermal, chemical, electrical, magnetic, etc.).

\section{Methods}
\label{Methods}

\subsection{Initial Filtering}
\label{Initial Filtering}

\begin{figure}
    \centering
    \includegraphics[width=1\linewidth]{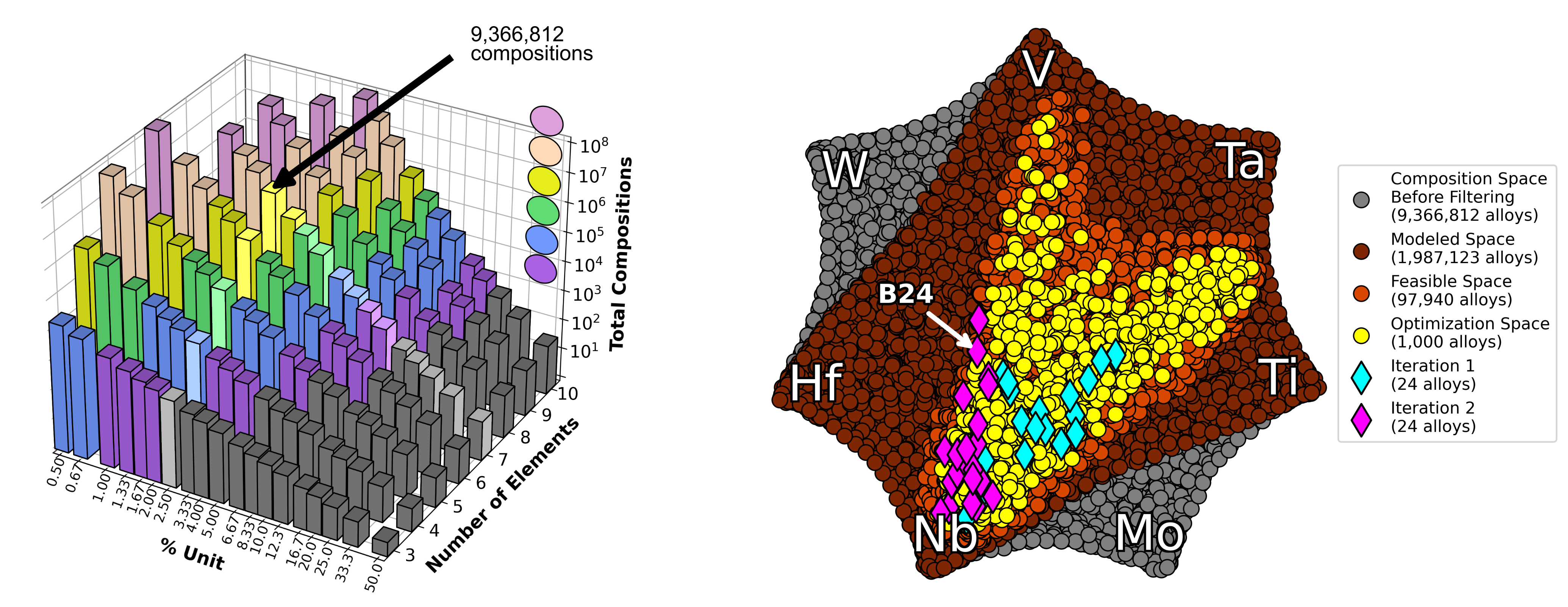}
    \caption{Alloy space size as a function of the number of alloying elements and compositional resolution (left), and a UMAP representation of the alloy space after each filtering step along with alloys selected for each iteration (right). B24 corresponds to the best alloy after two iterations, as discussed in section \ref{Objective Property Optimization Results}.}
    \label{fig:Alloy_Space}
\end{figure}

The compositional resolution for this work was chosen to be 2.5 at\% (i.e., each component could vary in 2.5 at\% increments). As shown in Figure \ref{fig:Alloy_Space} (left), there are $\sim$9.37 million candidate alloys at such a resolution in the 7-component Ti-V-Nb-Mo-Hf-Ta-W system. Figure \ref{fig:Alloy_Space} (right) shows the $\sim$9.37 million alloy space (“Composition Space Before Filtering”) using the Uniform Manifold Approximation and Projection for Dimension Reduction (UMAP) technique \cite{McInnes2018}, as well as the reduced spaces resulting from the various filtering and alloy selection steps discussed below. The UMAP is provided to help the reader visualize the sampling and distribution of the remaining alloys in the overall design space as the filtering steps discussed below were applied.

Exploring a space with over 9 million candidates is untenable for any experimental framework. To address this issue, several constraints were applied to intelligently filter the alloy space to a more tenable size. Before applying any constraints, the overall alloy system was first divided into “subsystems” corresponding to the particular combination of alloying elements in each composition. For example, all quaternary alloys containing some combination of Ti-V-Mo-W would comprise a subsystem, whereas all quinary alloys containing some combination Ti-V-Nb-Mo-W would comprise another. This was done to better track the alloy subsystems considered and provide an additional dimension for filtering and selecting candidate alloys. In the overall Ti-V-Nb-Mo-Hf-Ta-W system, there is one septenary, seven senary, 21 quinary, 35 quaternary, 35 ternary, and 21 binary subsystems, summing to a total of 120 unique subsystems. These subsystems were given numerical designations; the lone septenary subsystem was designated as subsystem “1”, the seven senary subsystems were designated “2” through “8”, and so on. 

Two preliminary constraints based on the authors’ experience and intuition were used to initially filter the candidate space. First, molybdenum and tungsten were each limited to $<$ 15 at\% due to the synthesis technique used in this work (i.e., vacuum arc melting, as discussed in section \ref{Vacuum Arc Melting}). The melting temperature of tungsten is very close to the boiling temperature of vanadium, making compositional control of vanadium difficult if large fractions of tungsten are present. Molybdenum content was limited to prevent the formation of volatile oxides during melting and homogenization. Second, the entire septenary subsystem was eliminated, as it was by far the largest and its alloys would also be the most time-consuming to synthesize. After these initial constraints were applied, high-throughput computational thermodynamic calculations were performed on the remaining $\sim$1.99 million compositions (“Modeled Space” in Figure \ref{fig:Alloy_Space}). Thermo-Calc software and its TCHEA6 database were used to predict equilibrium phase fractions and a range of material properties (e.g., density, solidus temperature, liquidus temperature) for each alloy. Custom Python scripts were utilized to automate these calculations using the Texas A\&M High-Performance Research Computing (HPRC) facility and Thermo-Calc’s Python API (TC-Python).

After the thermodynamic calculations were complete, the resulting equilibrium and property predictions were used to further filter the alloy space based on the reasons and corresponding constraints summarized in the following list:

\begin{itemize}
    \item Promote thermodynamic stability
    \begin{itemize}
        \item Ideal configurational entropy greater than or equal to equimolar ternary
        \begin{itemize}
            \item $S_\mathrm{conf}^\mathrm{ideal} \ge R \ln 3 = 9.134$ J/mol$\cdot$K
        \end{itemize}
        \item BCC single-phase at 800 °C
        \begin{itemize}
            \item Fraction BCC $>$ 99.9999\% to address rounding issues in Thermo-Calc
        \end{itemize}
    \end{itemize}
    \item Promote processability
    \begin{itemize}
        \item Narrow solidification range $\left( \Delta T = T_\mathrm{liquidus} - T_\mathrm{solidus} \right)$
        \begin{itemize}
            \item Lower half of all remaining alloys, which corresponded to $\Delta T < 264.5$ °C
        \end{itemize}
    \end{itemize}
    \item Promote material properties beneficial to a potential application
    \begin{itemize}
        \item Low density ($\rho$)
        \begin{itemize}
            \item Lower half of all remaining alloys, which corresponded to $\rho < 11.39$ $\mathrm{g/{cm}^3}$
        \end{itemize}
        \item High "melting point" $\left( T_\mathrm{m} = (T_\mathrm{solidus} + T_\mathrm{liquidus})/2 \right)$
        \begin{itemize}
            \item $T_\mathrm{m}$ was increased in 10 °C increments until the feasible space was reduced to $<$ 100,000 alloys, which corresponded to $T_\mathrm{m} >$ 2330 °C
        \end{itemize}
    \end{itemize}
\end{itemize}

As discussed in section \ref{Introduction}, increasing the configurational entropy of an alloy can promote thermodynamic stability against a variety of degradation mechanisms. Therefore, a threshold for ideal configurational entropy, calculated solely from composition using the Boltzmann entropy formula, was set to ensure that all feasible alloys had an ideal configurational entropy equal to or greater than that of an equimolar ternary alloy. 

The alloys were then filtered such that only those predicted by Thermo-Calc to be single-phase BCC remained. A single-phase solid solution of a particular composition will have a higher configurational entropy than a multi-phase counterpart. Furthermore, the current work did not consider microstructure (e.g., grain size and morphology) as an input variable in the optimization algorithm. As such, it was desired that every alloy had a similar microstructure to limit the error associated with ignoring these features. Finally, single-phase materials better lend themselves to predicting bulk properties via micromechanical testing (e.g., nanoindentation) than multi-phase materials. That is, localized testing of multi-phase materials requires additional consideration of the phase fractions sampled in each testing event (e.g., indent), which can add additional scatter and uncertainly to predictions of bulk properties.

A temperature of 800 °C was chosen for calculating the equilibrium phase fractions in Thermo-Calc. It was anticipated that using room temperature calculations would result in an excessively large number of alloys being eliminated by this constraint. Furthermore, the chosen alloying elements should all be sluggish solid-state diffusers below this temperature. Therefore, even if non-BCC equilibrium phases exist in a given alloy at lower temperatures, it would be unlikely that non-negligible fractions of such secondary phases could form via nucleation and growth during cooling from 800 °C. This work did not consider the formation of secondary phases via other transformation mechanisms (e.g., spinodal decomposition or diffusionless transformations).

Alloys with large solidification or “freezing” ranges (i.e., large differences between the solidus and liquidus temperatures) are known to be more susceptible to the formation of undesirable microstructural features and defects during melt-based processing, such as dendrite formation, alloying element segregation, or the formation of pores and cracks \cite{Piwonka1998}. Therefore, half of the remaining alloys with the largest solidification ranges were filtered from the design space to promote more desirable and homogeneous microstructures in the as-cast samples from vacuum arc melting.

Two final constraints were applied based on assumed application-based property requirements. As mentioned above, no specific application was envisioned for this work. However, most real-world applications will have specific requirements for material properties such as density and melting point. Therefore, it was desired to demonstrate the utility of such filtering methods in this work. To this end, the median density among the remaining alloys was taken as the maximum acceptable density, thereby cutting the remaining space in half again. Finally, the minimum acceptable “melting point” was increased in 10 °C increments until $\sim$100,000 alloys remained. In this work, the “melting point” was calculated as the average between solidus and liquidus temperatures predicted by Thermo-Calc. In hindsight, solidus temperature is perhaps a more meaningful constraint for high-temperature structural materials. However, it was subsequently determined that the resulting $\sim$100,000 alloy space is nearly identical whether solidus alone or the average between liquidus and solidus was used for this constraint.

The remaining $\sim$100,000 alloys were distributed among 40 subsystems. However, many of these remaining subsystems had exceedingly small alloy populations after the previous filtering steps were applied. Therefore, the 16 subsystems containing the smallest populations were eliminated, leaving the remaining feasible alloys distributed among 24 subsystems. This was done to align with the fact that alloys were to be synthesized and characterized in batches of 24 per iteration and that the numerical subsystem designations were used as a candidate selection criterion for the first iteration, as discussed in section \ref{Candidate Selection}. This reduced the overall space slightly to 97,940 alloys (“Feasible Space” in Figure \ref{fig:Alloy_Space}). 

\begin{figure}
    \centering
    \includegraphics[width=0.5\linewidth]{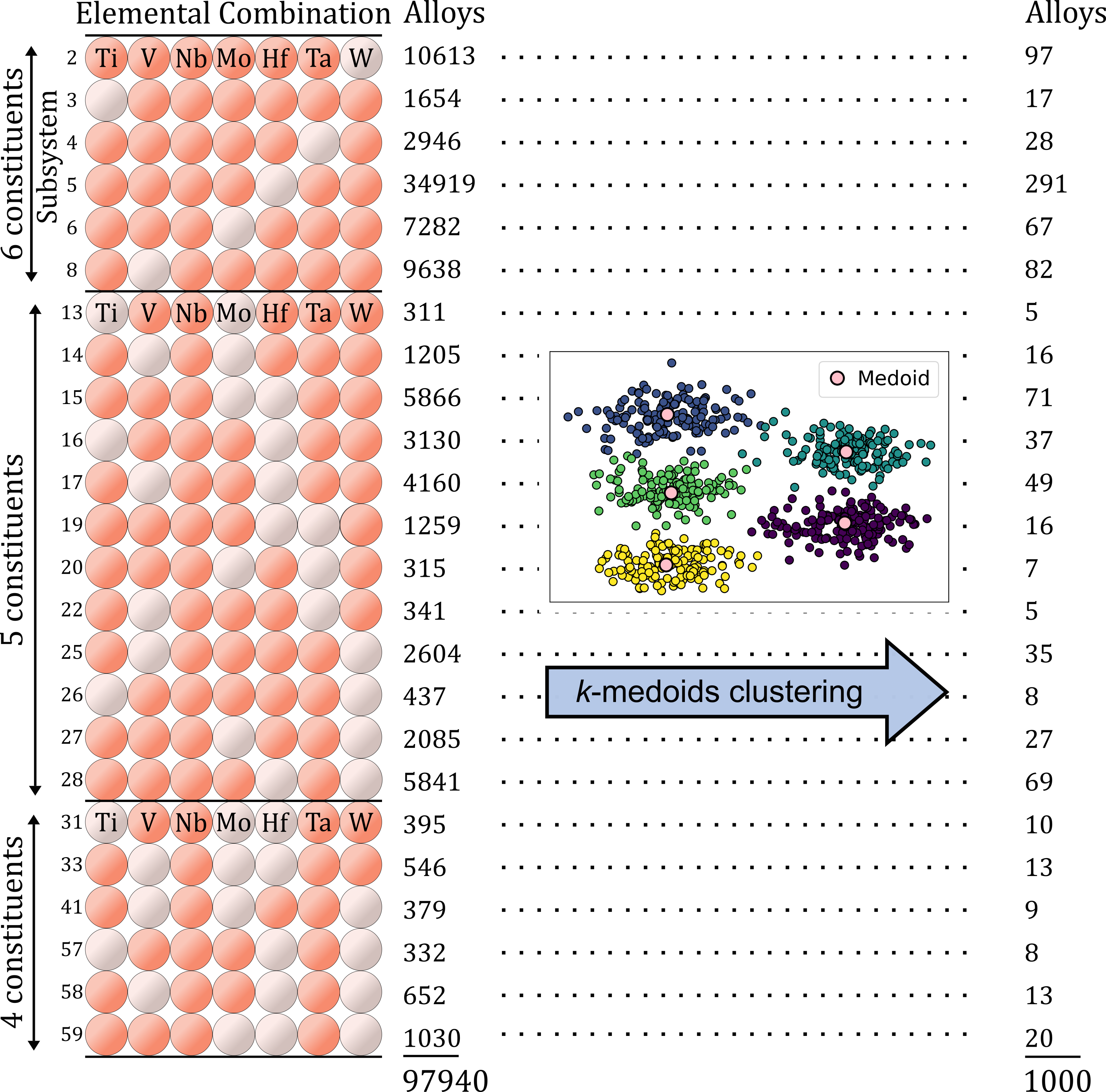}
    \caption{Subsystem types and their relative sizes in the "Feasible Space" (left) and the further reduced "Optimization Space" (right) selected via k-medoids clustering. The circles represent whether or not a particular alloying element is present in a particular subsystem (dark = yes, light = no).}
    \label{fig:k-medoids}
\end{figure}

A final filtering step was applied to further reduce the design space while ensuring that the resulting set provided a representative sampling of the entire feasible space of 97,940 alloys. Ideally, a representative sample would contain a subset of alloys uniformly distributed over the feasible space. While this would be trivial for binary or ternary systems, it becomes increasingly non-trivial for higher dimensional compositional spaces. To achieve this goal, 1,000 alloys nearly equally distributed over the feasible space were selected using a $k$-medoids clustering algorithm \cite{Park2009}. The $k$-medoids method is particularly effective in identifying representative elements in complex, multidimensional data spaces. The technique partitions data into a predefined number of groups or clusters based on similarity, choosing a central point or “medoid” within each cluster that minimizes the distance to all other points in the cluster. In this work, the similarity between alloys was determined by the Euclidean distance between their compositions, incorporating both the atomic fractions and the numerical designation of each alloy’s subsystem to ensure that a medoid would be calculated for each of the 24 subsystems while retaining a minimum distance between points across the overall compositional space. By this method, the feasible space was partitioned into 1,000 clusters, and the corresponding medoids were the 1,000 alloys selected. The results of this clustering process and the size of the remaining 24 subsystems are shown in Figure \ref{fig:k-medoids}. This set of 1,000 alloys corresponds to the “Optimization Space” in Figure \ref{fig:Alloy_Space}, which constituted the down-selected alloy space from which all candidates were selected for synthesis and characterization in each iteration. 

\subsection{Candidate Selection}
\label{Candidate Selection}

No prior relationship between composition and mechanical properties was assumed. Therefore, the alloy chemistries tested in the first iteration could not be selected based on any prior knowledge or predictions of relationships between inputs (i.e., composition) and the objective properties. The simplest way to address this lack of prior knowledge would be to select alloys for the first iteration purely at random from the down-selected 1,000-alloy candidate space. However, such a method would run the risk of two or more alloys being coincidentally grouped in the same region of the compositional space, thereby limiting mechanical property vs. composition information provided to the BBO algorithm. Of course, one could incorporate mechanical property vs. composition predictions using physics-based models. However, such modeling is computationally expensive and runs the risk of incorporating false biases into the optimization algorithm. Therefore, the approach chosen in this work was to select alloys with compositions intentionally dispersed throughout the candidate space to ensure representative sampling without significant compositional overlaps. To this end, the $k$-medoids clustering algorithm discussed in section \ref{Initial Filtering} was used to select 24 alloys by splitting the 1,000-alloy “Optimization Space” into 24 clusters and selecting the corresponding medoid of each cluster. This was done to ensure that the selected alloys were widely distributed in the compositional space while also ensuring that one alloy was selected from each of the 24 possible subsystems.

As the title of this manuscript suggests, only two iterations were produced and characterized in this work. The 24 alloys produced in the first iteration (“Iteration 1” in Figure \ref{fig:Alloy_Space}) were named A01 through A24. The 24 alloys constituting the second iteration (“Iteration 2” in Figure \ref{fig:Alloy_Space}) were selected via the optimization techniques described in section \ref{Batch Bayesian Optimization}, which was informed by the data collected from the methods described in section \ref{Characterization}. The second iteration alloys were named B01 through B24. The nominal compositions, naming convention, and measured objective properties of all alloys produced in this project are given in the supplemental data. 

\subsection{Synthesis and Processing}
\label{Synthesis and Processing}

\subsubsection{Vacuum Arc Melting}
\label{Vacuum Arc Melting}

All 48 alloys among both iterations were synthesized from high-purity elements ($>$ 99.9 wt\%) using a Beuhler AM200 vacuum arc melter (VAM). VAM was performed under a highly purified argon atmosphere, achieved by evacuating the chamber to $<$ 0.01 Pa and refilling it with ultra-high purity argon (Linde Gas, $>$ 99.999\%) to 80 kPa three times. Titanium sponge was then melted in the chamber to serve as a getter for any remaining oxygen or nitrogen gas. To enable high-throughput synthesis, a custom crucible was deployed with nine sample slots for concurrent batch-wise melting of “buttons” of each alloy – each button weighing $\sim$8 g and measuring $\sim$10 $\times$ $\sim$5 $\times$ $\sim$18 mm. Each alloy was re-melted a minimum of five times, flipping the button between each melt to promote full melting of the entire sample and homogeneous distribution of the alloying elements.

\subsubsection{Homogenization Heat Treatment}
\label{Homogenization Heat Treatment}

After VAM, heat treatments were performed on all samples to further homogenize the alloying elements and eliminate any as-cast dendrites. These treatments were performed using an all-metal furnace (Centorr LF Series, Model 22) under an inert atmosphere. The inert atmosphere was produced by three cycles of evacuating the furnace to $<$ 0.01 Pa and refilling to atmospheric pressure with ultra-high purity argon (Linde Gas, $>$ 99.999\%). Heat treatments were performed at 1925 °C for 24 hours followed by furnace cooling to room temperature. This thermal profile was determined from predictions of the temperature and time required for all alloying elements to diffuse across the dendrite to the inter-dendritic region and, therefore, fully homogenize the composition. This thermal profile was predicted via Thermo-Calc’s diffusion module (DICTRA) using methods previously published by the authors \cite{Vela2023}.

\subsection{Characterization}
\label{Characterization}

To enable characterization, each button was cut via wire electrical discharge machining (wire-EDM) into elliptical cross-sectional discs from the center of each button. Each of these cross-sectional discs had dimensions of $\sim$10 $\times$ $\sim$5 $\times$ $\sim$3 mm. These discs were then used for polishing, scanning electron microscopy (SEM), energy dispersive spectroscopy (EDS), X-ray diffraction (XRD), Vickers microhardness, and nanoindentation. 

\subsubsection{Metallography and Microscopy}
\label{Metallography and Microscopy}

Due to the large number of unique alloys being produced and the varying properties among them, it was necessary to develop a custom metallographic preparation technique that was suitable for all alloys and all characterization methods. To this end, each sample was first mounted in conductive thermosetting resin using an Allied High Tech TechPress 3. The mounted samples were ground using 15 µm then 9 µm diamond grinding discs, followed by polishing using 9 µm then 3 µm polycrystalline diamond suspensions. The final step used an attack polishing solution consisting of 3 parts 0.04 µm colloidal silica to 1 part 30\% hydrogen peroxide (VWR). All grinding/polishing steps were performed using an Allied High Tech MetPrep 4 with a 12-sample holder. All metallography consumables were purchased from Allied High Tech.

After polishing, microstructural analysis was performed using an SEM (Thermo Fisher Phenom XL). Micrographs were acquired at 500X, 1000X, and 2000X using a backscattered electron detector (BSD), a 15 kV accelerating potential, and a $\sim$6 mm working distance. BSD was chosen for its ability to highlight grain structure and microstructural homogeneity without the need for chemical etching.

\subsubsection{Composition and Phase Analysis}
\label{Composition and Phase Analysis}

As the only inputs for the objective function being optimized, reliable compositional data was paramount in this work. Furthermore, the large number of alloys produced in such a project necessitates the utilization of high-throughput chemical analysis techniques. Therefore, great care was taken to identify a chemical characterization technique that could produce both rapid and reliable results.

\begin{figure}
    \centering
    \includegraphics[width=0.5\linewidth]{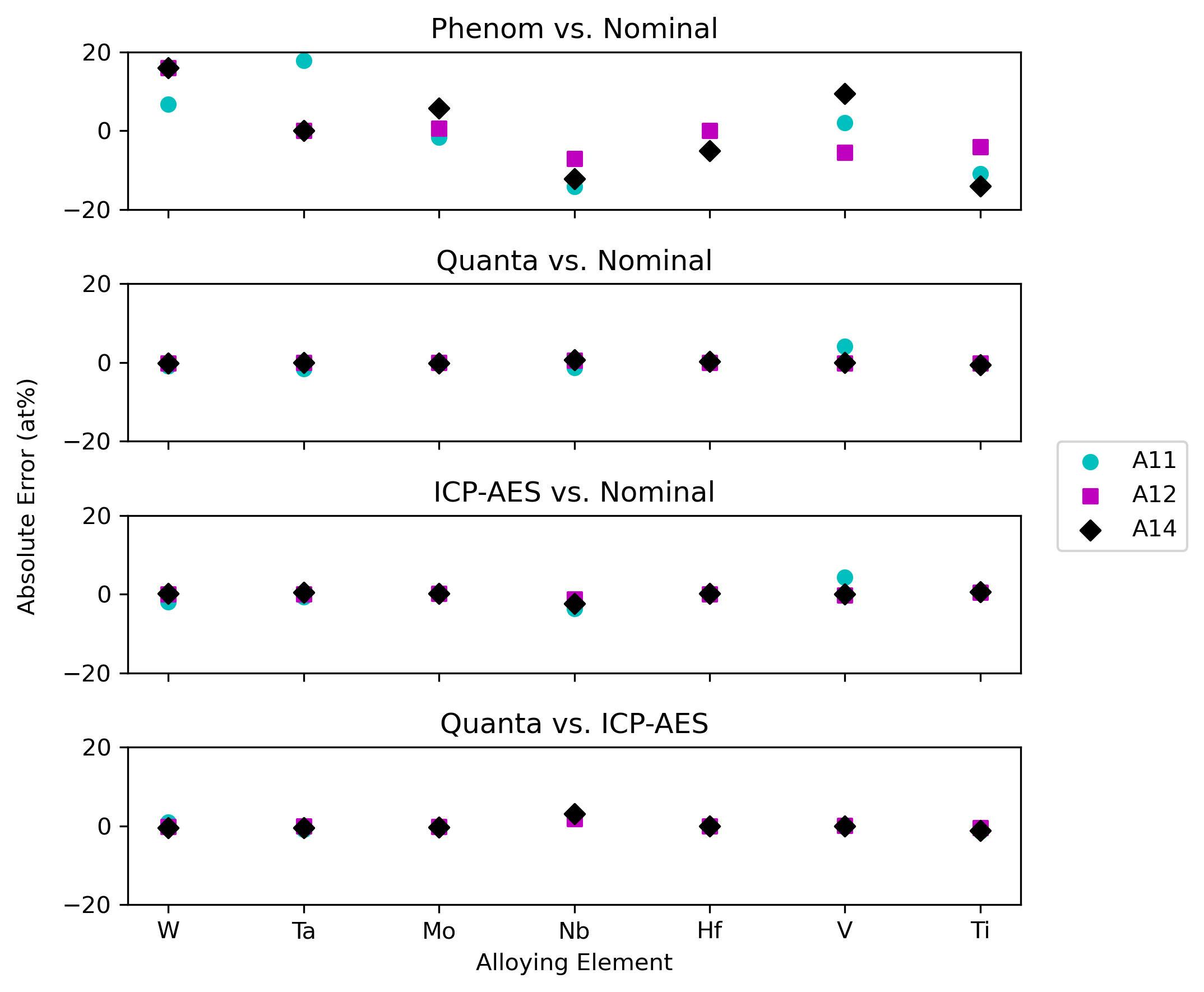}
    \caption{Comparison of chemical analysis results from two different EDS systems and ICP-AES vs. the nominal compositions of three alloys produced in the first iteration.}
    \label{fig:EDSvsICP}
\end{figure}

Chemical analyses of all alloys produced in the first iteration were carried out using EDS detectors on two different SEM systems: a Thermo Fisher Phenom XL (“Phenom”) and an FEI Quanta 600 (“Quanta”). The Phenom data exhibited significant deviations from the nominal compositions, whereas the Quanta data yielded results more closely aligned with the nominal compositions. Therefore, to better assess and compare the reliability of each EDS system, three alloys from the first iteration (A11, A12, and A14) underwent additional chemical analysis via inductively coupled plasma atomic emission spectroscopy (ICP-AES). ICP-AES is well-known to be a very reliable method for obtaining quantitative fractions of elements in metal alloys. However, the process is relatively cumbersome and would be impractical for measuring the compositions of all alloys in a high-throughput alloy discovery project. Therefore, the implementation of the ICP-AES was to test and verify the more agile EDS technique. The ICP-AES results corroborated the reliability of the Quanta’s EDS measurements while confirming the discrepancies in the Phenom’s data. It is speculated that the Phenom’s EDS detector lacks the necessary X-ray energy resolution to distinguish between peaks in the EDS spectra of certain elements (i.e., Ta/W and Nb/Mo), leading to substantial error in the resulting compositional data. The results of this comparison between the chemical characterization techniques and equipment are summarized in Figure \ref{fig:EDSvsICP}.

After verifying its reliability, EDS via the Quanta was chosen as the primary method for assessing alloy composition and chemical homogeneity in all alloys produced. These measurements were conducted using a 20 kV accelerating potential and an 11 mm working distance with an Oxford Instruments EDS detector. At least two low-magnification area scans from unique regions were taken on each sample to assess the overall composition. Furthermore, at least eight point-wise measurements were taken on each sample to verify the uniformity of composition and dissolution of dendrites during homogenization.

As discussed in section \ref{Initial Filtering}, a primary constraint in this work was that all alloys were single-phase BCC. To assess whether this constraint was met and, therefore, the thermodynamic predictions were reliable, phase analysis was performed on all alloys using a Bruker D8 Discover X-ray diffractometer equipped with a Cu-K\textsubscript{$\upalpha$} X-ray source and a Vantec 500 area detector. After each scan, the background was subtracted using algorithms provided in Bruker’s analysis software. 

\subsubsection{Nanoindentation}
\label{Nanoindentation}

The results of the nanoindentation experiments served as the primary means of assessing the objective properties to be optimized in this study. Furthermore, as a microscale and highly localized mechanical characterization method, nanoindentation is prone to significant error unless proper care is taken when preparing the samples, taking the measurements, and analyzing the data. Therefore, a systematic approach was undertaken to maximize the veracity of the data acquired. To this end, sample criteria and testing parameters followed the guidelines specified in the ISO 14577 standard \cite{ISO2015}. The specific steps taken to ensure the reliability of the nanoindentation data are detailed in the experimental decision tree provided in the supplemental data.

Nanoindentation was performed with a diamond Berkovich indenter using a Nanomechanics iMicro2. Before use, indenter tips were examined with a laser-optical profilometer (Keyence VK-X1000 3D Surface Profiler). If deemed necessary, tips were cleaned and re-imaged before use. Tip area functions (i.e., equations used to calculate projected contact area based on indentation depth) \cite{Fischer-Cripps2011} were calibrated by indentation of a fused silica standard (Edmunds Optics, Stock No. 45-309). Once samples were loaded for indentation, ambient testing conditions were verified to meet those outlined by the ISO 14577 standard \cite{ISO2015}. Because nanoindentation performs displacement measurements on the nm scale, it is prone to appreciable error due to environmental vibrations or thermal expansion or contraction of the indenter equipment and/or the sample after loading, often referred to as “thermal drift”. To minimize error associated with such drift, indentation was not initiated until at least 30 minutes had passed after loading the sample or a measured drift rate of $\leq$ 0.05 nm/s was verified.

To ensure the comparability of indentation data collected across a wide range of sample properties, machine-level indentation parameters were maintained across all samples. The parameter values were determined using an expected range of sample properties within the selected alloy space based on the authors’ previous experience with refractory metals. The specific values for parameters used throughout this study are given in the supplemental data. The continuous stiffness method (CSM) \cite{Oliver1992} was used to capture detailed measurements of the indentation response throughout the loading duration. To reduce the influence of the indentation size effect \cite{Fischer-Cripps2011,Pharr2010,Stelmashenko1993,Nix1998} on measured properties, the largest indentation depth that could be reliably obtained across the alloy space was estimated at 2 µm and used as the target depth for all tests. An indentation strain rate of 0.2 $\mathrm{s}^{-1}$ was used throughout, as it provides low indentation times, is well within the working range of CSM for most materials, and largely avoids sources of error introduced by excessively high or low indentation rates \cite{Fischer-Cripps2011}.

At least ten indents were made on each sample, which were placed in a square 3 $\times$ 3 array with the tenth indent used as a fiducial marker to enable the subsequent image analysis techniques discussed in section \ref{Procedures for Pile-up Correction}. A depiction of the indentation array geometry is shown in the supplemental data. A center-to-center indent spacing of 100 µm was used to prevent interaction between the plastic zones around each indent. 

The basic definitions \cite{Fischer-Cripps2011} of indentation hardness ($H$), reduced modulus ($E_\mathrm{r}$), and sample modulus ($E_\mathrm{s}$) are defined in equations (\ref{eq:H=}), (\ref{eq:E_r=}), and (\ref{eq:E_s=}), respectively. Sample hardness is simply the applied load ($P$) divided by the projected contact area ($A_\mathrm{c}$); the projected contact area being the contact surface of an indent projected orthogonally to the applied load. When determining sample modulus, the projected contact area and the stiffness of contact $(S)$ can be used to determine the reduced modulus $(E_\mathrm{r})$. However, converting this value to the sample modulus $(E_\mathrm{s})$ requires additional information, as shown in equation (\ref{eq:E_s=}). To enable this calculation, the indenter’s modulus $(E_\mathrm{i}=1141\ \text{GPa})$ and Poisson’s ratio $(\nu_\mathrm{i}=0.07)$ were assumed using widely accepted values for diamond. The Poisson’s ratio of an arbitrary metal is typically very close to 0.3. Therefore, $\nu_\mathrm{s}=0.3$ was assumed for every alloy tested. 

\begin{equation}
\label{eq:H=}
H = \frac{P}{A_\mathrm{c}}
\end{equation}

\begin{equation}
\label{eq:E_r=}
E_\mathrm{r} = \frac{\sqrt{\pi}}{2} \frac{S}{A_\mathrm{c}}
\end{equation}

\begin{equation}
\label{eq:E_s=}
E_\mathrm{s} = \frac{1 - {\nu_\mathrm{s}}^2}{\frac{1}{E_\mathrm{r}}-\frac{1-{\nu_\mathrm{i}}^2}{E_\mathrm{i}}}
\end{equation}

\subsubsection{Procedures for Pile-up Correction}
\label{Procedures for Pile-up Correction}

\begin{figure}
    \centering
    \includegraphics[width=0.5\linewidth]{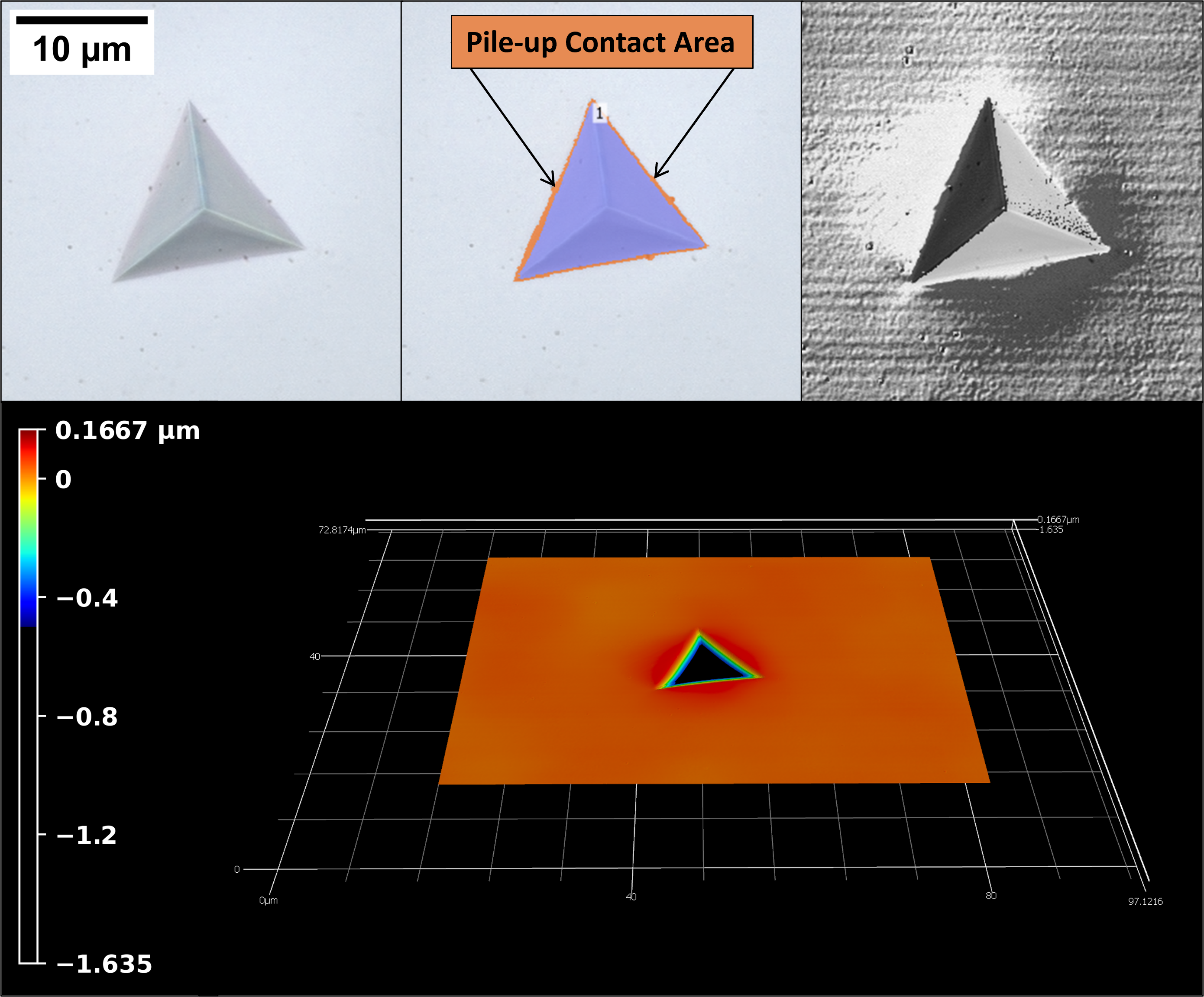}
    \caption{Optical micrograph of a representative indent (top-left). Visual demonstration of how the projected contact area was corrected for pile-up using profilometry data; purple corresponds to the uncorrected projected contact area, and orange corresponds to the additional projected contact area due to pile-up (top-center). Differential interference contrast (DIC) micrograph of the indent (top-right). 3D topography map used to directly measure additional contact area due to pile-up (bottom).}
    \label{fig:Pile-up}
\end{figure}

Although CSM utilizes dynamic load oscillation to enable the continuous measurement of the material’s contact stiffness, it is unable to account for the contact area contributions of “pile-up” commonly exhibited by metal alloys during nanoindentation \cite{Fischer-Cripps2011,Bolshakov1998}. If left uncorrected, this change in contact area can produce significant error in the hardness and modulus measurements.

To correct for the effect of pile-up, high-resolution scans were taken of each indent using the Keyence profilometer described in section \ref{Nanoindentation}. Each indent’s true projected contact area was then determined using the resulting topography maps. This method is demonstrated visually in Figure \ref{fig:Pile-up}. The regular indentation array and fiducial marking described in section \ref{Nanoindentation} allowed for this process to be automated through the profilometer’s software. Corrected hardness was determined by equation (\ref{eq:H=}) using the true projected contact area as $A_\mathrm{c}$ and the maximum load applied during indentation as $P$. Corrected modulus was determined by equations (\ref{eq:E_r=}) and (\ref{eq:E_s=}) using the true projected contact area as $A_\mathrm{c}$ and the stiffness value calculated from the unloading portion of the indentation curve as $S$.

\subsubsection{Vickers Microhardness}
\label{Vickers Microhardness}

Additional hardness measurements using a larger scale technique were desired for comparison with the nanoindentation hardness data. To achieve this, Vickers microhardness measurements were taken using a LECO LM-100 Microhardness Tester with an applied load of 4.90 N (500 gf) and a dwell time of 10 s. Five Vickers indents were made across each sample. The minimum distance between indents and between any indent and the sample edge was verified to be at least five times the maximum diagonal of any indent per the ASTM E92-17 standard \cite{ASTM2017}.

\subsection{Batch Bayesian Optimization}
\label{Batch Bayesian Optimization}

Bayesian optimization (BO) is an advanced computational technique that applies Bayes' theorem across all design strategies. The theorem posits that predictions based on known data are inherently more reliable than those without such backing. Such data can be derived from prior experiments, literature values, or any other relevant source. Furthermore, as additional data points are introduced, predictions should be dynamically updated to reflect the new information. Therefore, the technique can be applied to discover where an optimum design region exists within unexplored or unknown $n$-dimensional Euclidean spaces via the collection of new observations \cite{Jones1998,Greenhill2020,Arroyave2022}. As mentioned above, no prior knowledge of how the inputs (chemical composition) related to objective properties (specific hardness and modulus) was assumed in this work. Therefore, the application of prior knowledge to inform predictions of new alloys with improved properties was only possible after the first iteration of samples were produced and characterized. Only two iterations of alloys were produced in this work, meaning only a single round of optimization was conducted. As such, this study served as a test of how quickly such a method can produce significant improvements in objective properties starting from an essentially blank slate.

BO requires the selection of a method for evaluating prior knowledge to inform future predictions. The method chosen for this study was a “Gaussian process” (GP) regressor, which acts as a probabilistic model over a range of potential solutions to the objective problem(s) \cite{Bradford2018,Khatamsaz2021}. Real-world design problems are often limited to inputs and outputs with no previously known relations composed of a finite combination of algebraic operations, giving them a “black box” form. One of the significant advantages of a GP is its intrinsic ability to recognize its own information deficits, providing a measure of model uncertainty. This quality sets it apart from other predictive methods like neural networks or random forests, which may not inherently account for their own limitations in knowledge, thereby leading to erroneous predictions with seemingly inexplicable confidence.

A GP predicts the mean of a black box function’s response to the inputs, including a variance at each position in the input space using a kernel function chosen for performing such predictions. In this work, we used the squared exponential covariance kernel function described in \cite{Khatamsaz2021}. The GP method incorporates a distance-dependent correlation metric, encoding how similar function outputs are expected to be based on the proximity of their inputs. Consequently, the variance of a predicted value typically increases as the input moves further away from the data points in the training set, reflecting greater uncertainty in regions with less data. By inputting material composition and property data into a GP and setting a specific output objective (e.g., specific hardness or specific modulus), the GP will provide information for how the inputs are related to the output as well as its own uncertainty in that information at each position in the input space. In this work, two outputs (i.e., objective properties) were considered. Therefore, a separate GP model was required for each objective property.

The last step in a single iteration of the BO process requires the use of an “acquisition function”, which is used to guide future input selections by accounting for both the mean predicted response and variance calculated by the GPs. An acquisition function balances two key criteria: firstly, selecting new candidates that will provide the most novel information, and secondly, choosing those predicted to surpass the performance of the currently best-known candidates based on the previous data. In other words, it balances the exploitation of current knowledge and exploration of unknown space using the variance supplied by the GP based on its own uncertainty. The acquisition function used in this work was Expected Hypervolume Improvement (EHVI), a function shown to be successful in multiple environments for higher dimensional optimization problems \cite{Emmerich2011,Zhao2018}.

As done in this work, BO can be executed in batches (i.e., batch Bayesian optimization or BBO) \cite{Couperthwaite2020}. The physical instruments behind experimentation (e.g., material synthesis/processing, microstructural characterization, mechanical testing) are typically much more efficient when such experiments can be done in parallel. Therefore, the primary advantage of the BBO method in experimental materials discovery is its efficiency. Synthesizing and characterizing materials sequentially (i.e., turning sequential Bayesian optimization into an experimental materials discovery strategy) would be needlessly time-consuming and resource-intensive by comparison.

As with many similar computational methods, the GPs used in this work require a selection of hyperparameters (e.g., scaling factor and length scale, described below). Furthermore, the hyperparameter selections can have a profound impact on the efficiency of the optimization process. However, no prior knowledge of how hyperparameters affected GP efficiency was assumed in this work. Instead, hundreds of GPs were run for each objective property, thereby providing a statistical sampling of the results over a wide range of hyperparameters. This method avoids the process of identifying ideal hyperparameters, at the cost of extra computational power required to run hundreds of iterations of BO. Of course, fewer BO iterations would be required if the hyperparameters were initially optimized. However, that process may be non-trivial, could require substantial human effort upfront, and could still result in non-ideal hyperparameters. Furthermore, each BO calculation is computationally cheap, particularly when the candidate space has been filtered to only 1000 alloys. In fact, all iterations of BO calculations required in this project were completed in a manner of minutes on a laptop computer.

The scaling factor tells the GP how dramatic it should assume its uncertainty in a prediction of an objective property is for a given set of inputs. In this work, the scaling factor of each GP was set as a constant based on the range of observed values for the corresponding objective property in the first iteration. Specifically, the normalized difference between the “worst” and “best” sample in the first iteration was increased by 20\% and used to indicate how far the model should assume a true objective property value could deviate from the calculated mean for a given set of input values. As two GPs were run simultaneously (one for each objective property), two scaling factors (one for each GP) were defined using data from the first iteration and used for all predictions for the second iteration.

The length scale tells the GP how sensitive it should assume the objective properties are to each input (i.e., how “wiggly” the objective property function is with respect to each input). Being that the inputs were composition, the GP’s length scales corresponded to molar fractions of each element. Therefore, if the length scale was 0.05, the GP would assume that significant changes in the objective function are possible over a length scale of 5 at\% for that particular alloying element. There were seven elements possible in each candidate alloy, meaning seven length scales needed to be defined for each GP. Each of the seven necessary length scales was selected from a possible range of 0.05 to 0.95 (i.e., 5 to 95 at\%) using Latin Hypercube Sampling. Therefore, length scales $<$ 0.05 (nearly all variance) and $>$ 0.95 (nearly no variance) were forbidden. This provided a near-random sampling of length scales over a wide and reasonable range for each input in the respective GPs.

Using the methods described above for selecting hyperparameters, the two GPs were run simultaneously using the data from the first iteration of alloys. The results of both GPs were then simultaneously passed to the EHVI function, which identified one alloy from the 976 remaining in the “Optimization Space” (see section \ref{Initial Filtering}) calculated to have the highest probability of increasing one or both of the objective properties (i.e., the alloy with the best chance of expanding the hypervolume). This process was then repeated 500 times, resulting in a list of 500 recommendations for alloys that had the best chance of expanding the hypervolume. As would be assumed, the list of 500 alloys contained many duplicates. After all duplicates were removed, 60 unique alloy compositions remained. From this list of 60 unique alloys, an additional round of $k$-medoids clustering was used to select 24 alloys to be produced and characterized in the second iteration.

One could argue that a better method for selecting new alloys for the second iteration would be based on their plurality in the overall list of 500 recommended alloys. For example, if a particular alloy composition was identified 300 times among the 500 BO calculations, that alloy is calculated to have an excellent probability of expanding the hypervolume over a wide range of unique hyperparameter sets. Therefore, one could suggest that the 60 unique alloys should be rank-ordered based on plurality in the list of 500, and the top 24 alloys in that list of 60 should be selected for the second iteration. However, the initial 24 alloys from the first iteration could have provided a bias towards a local maximum in the objective properties. Therefore, such a method runs an increased risk of ignoring the region in the input-output hyperspace with the global maximum. 

\section{Results and Discussion}
\label{Results and Discussion}

\subsection{Chemistry and Phase Verification}
\label{Chemistry and Phase Verification}

\begin{figure}
    \centering
    \includegraphics[width=1\linewidth]{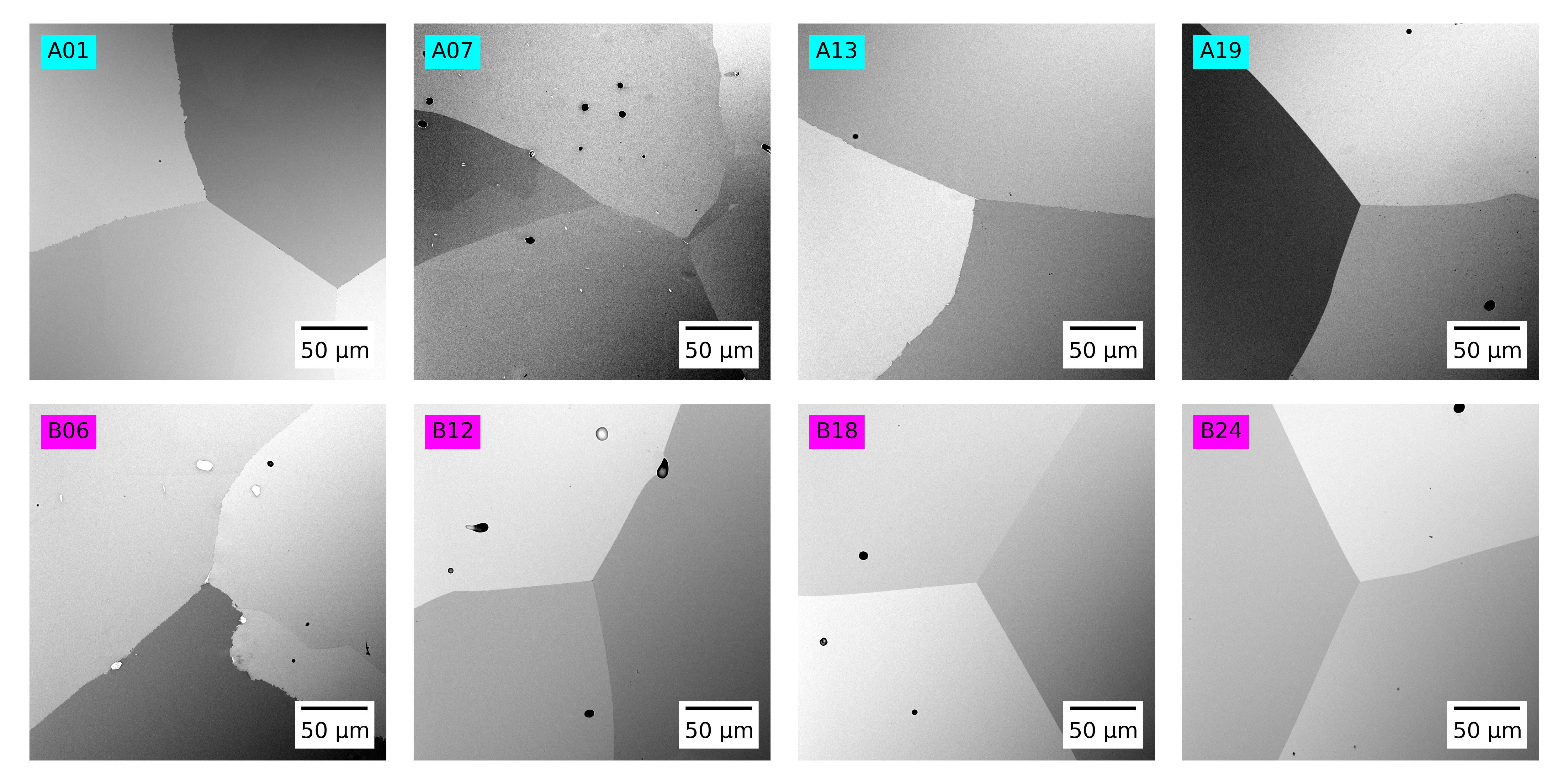}
    \caption{Backscattered electron (SEM-BSD) micrographs of representative samples from both iterations. Compositions for each alloy are given in the supplemental data. Dark contrast spots are pores. Light contrast spots in A07 and B06 are believed to be HfO\textsubscript{2} precipitates, as indicated by EDS analysis.}
    \label{fig:Micrographs}
\end{figure}

Representative SEM micrographs of eight alloys (four from each iteration) are shown in Figure \ref{fig:Micrographs}. The micrographs of all alloys indicate that homogenization was successful at removing the dendritic microstructure formed during arc melting. Residual porosity remains within the material due to the manufacturing processes employed in this high-throughput study. Additionally, several of the hafnium-containing alloys showed white precipitates (e.g., A07 and B06 in Figure \ref{fig:Micrographs}). Among the alloying elements, hafnium has the highest affinity for oxygen \cite{Backman2019}. Therefore, hafnium will serve as an oxygen getter, resulting in the formation of HfO\textsubscript{2} precipitates during alloy synthesis. To confirm this assumption, EDS line scans were performed across many of these precipitates in several alloys. These line scans showed clear spikes in both hafnium and oxygen concentrations at regions corresponding to the precipitates. As shown in the micrographs, both the pores and precipitates are only a few µm across and constitute a negligible fraction of each alloy’s total volume. Therefore, it was concluded that these defects should have a minor effect on the characterization methods employed in this study. As such, attempts to further reduce the size or population of such defects were considered to be beyond the scope of this work.

\begin{figure}
    \centering
    \includegraphics[width=0.5\linewidth]{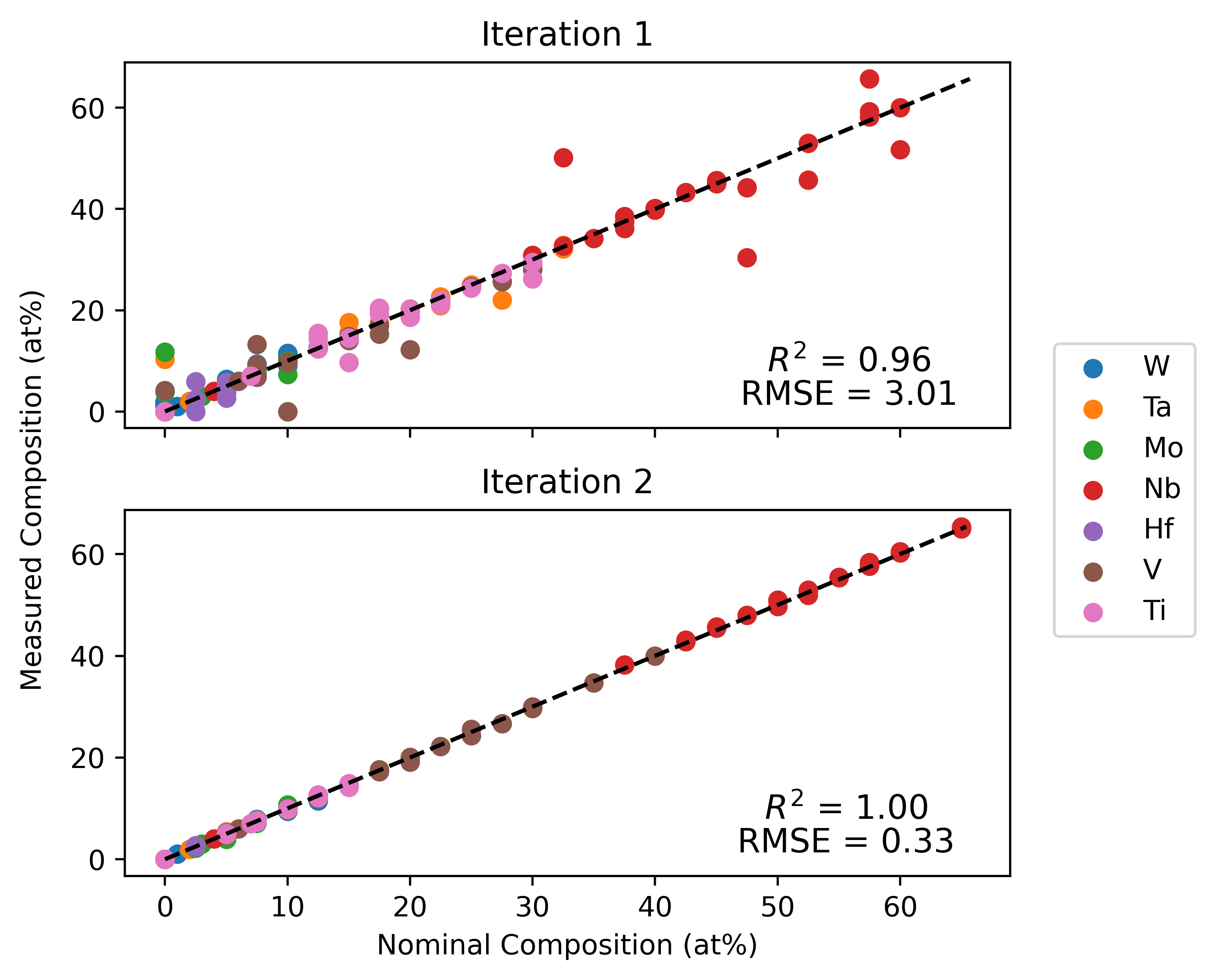}
    \caption{Parity plots of measured and nominal compositions by alloying element in each iteration.}
    \label{fig:Composition_Parity}
\end{figure}

The parity plots shown in Figure \ref{fig:Composition_Parity} illustrate the relative success in achieving target compositions across the two iterations. The $R^2$ (coefficient of determination) and RMSE (root mean square error) values serve as quantitative indicators of the extent to which the measured atomic fractions $(X_{\mathrm{measured},i})$ correspond with the nominal/target atomic fractions $(X_{\mathrm{nominal},i})$ among the $n$ possible components. Both $R^2$ and RMSE were calculated from an assumed perfect correlation (i.e., an assumed linear fit of $y = x$) using equations (\ref{eq:r-squared}) and (\ref{eq:RMSE}). 

\begin{equation}
\label{eq:r-squared}
    R^2 = 1 - \frac{\sum_{i=1}^n \left( X_{\mathrm{measured},i}-X_{\mathrm{nominal},i} \right)^2}{\sum_{i=1}^n \left( X_{\mathrm{measured},i}-\overline{X_\mathrm{measured}} \right)^2}
\end{equation}

\begin{equation}
\label{eq:RMSE}
    \mathrm{RMSE} = \sqrt{\frac{1}{n} \sum_{i=1}^n \left( X_{\mathrm{measured},i}-X_{\mathrm{nominal},i} \right)^2}
\end{equation}

Compositionally complex refractory alloys can be particularly difficult to synthesize due to the large ranges in melting points and vapor pressures among the alloying elements. For this reason, fairly significant error was observed for some elements in some alloys in the first iteration (i.e., $\mathrm{RMSE} = 3.01$). While such experimental error could become an issue for later iterations in a design cycle, it was not considered to be particularly problematic for the first iteration. As discussed in section \ref{Candidate Selection}, the first iteration alloys were simply selected using $k$-medoids clustering and without any prior knowledge of how composition related to the objective properties. Furthermore, the \textit{measured} compositions were used to inform the selection of alloys in the second iteration, effectively eliminating any consequences of such error. Notably, the second iteration demonstrated excellent alignment with target compositions (i.e., $\mathrm{RMSE} = 0.33$). This improvement is attributed to refinements in the synthesis and processing methods discussed in section \ref{Vacuum Arc Melting}.

\begin{figure}
    \centering
    \includegraphics[width=1\linewidth]{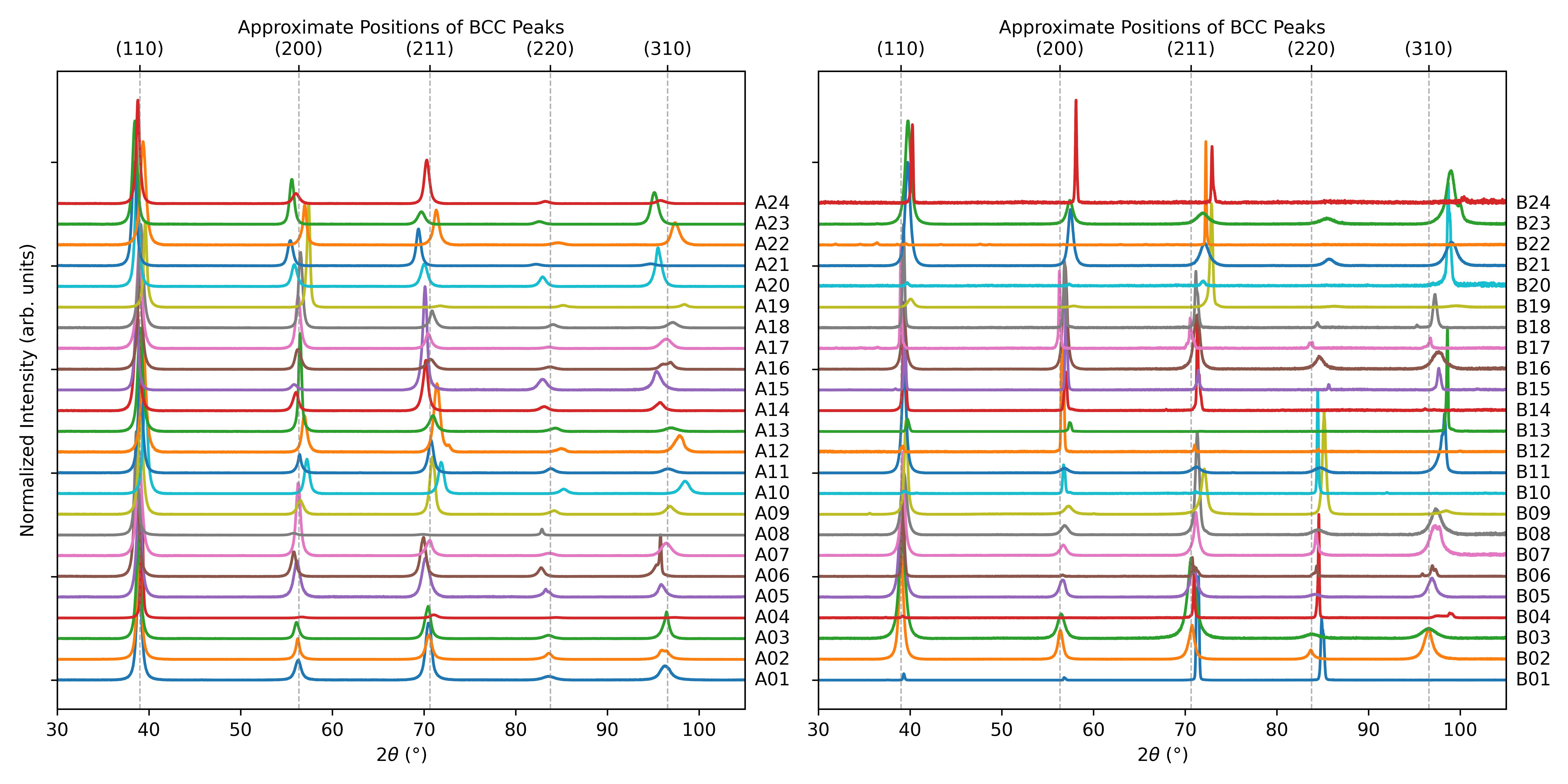}
    \caption{XRD spectra for all samples from the first (left) and second (right) iterations.}
    \label{fig:XRD}
\end{figure}

As shown in Figure \ref{fig:XRD}, results from the XRD analysis indicate that a BCC phase is present and dominate in all the alloys among both iterations. Minor non-BCC peaks are visible on some of the spectra (e.g., B06, B09, B14, B17, B18, B22, B24). It was speculated that these additional peaks could be the result of BCC superlattice reflections due to some degree of ordering in the microstructure \cite{Chen2019}. However, the peak positions do not lie at the corresponding $2 \theta$ values for such reflections. Regardless, the fact that the secondary peaks are very small indicates that the vast majority of the microstructure is single-phase BCC, as designed. This is a strong indication that the Thermo-Calc predictions are reliable. Dealing with minor fractions of such phases could become important for manufacturability and more applied testing, but either of those considerations are beyond the scope of this work. 

\subsection{Indentation Data}
\label{Indentation Data}

\begin{figure}
    \centering
    \includegraphics[width=0.5\linewidth]{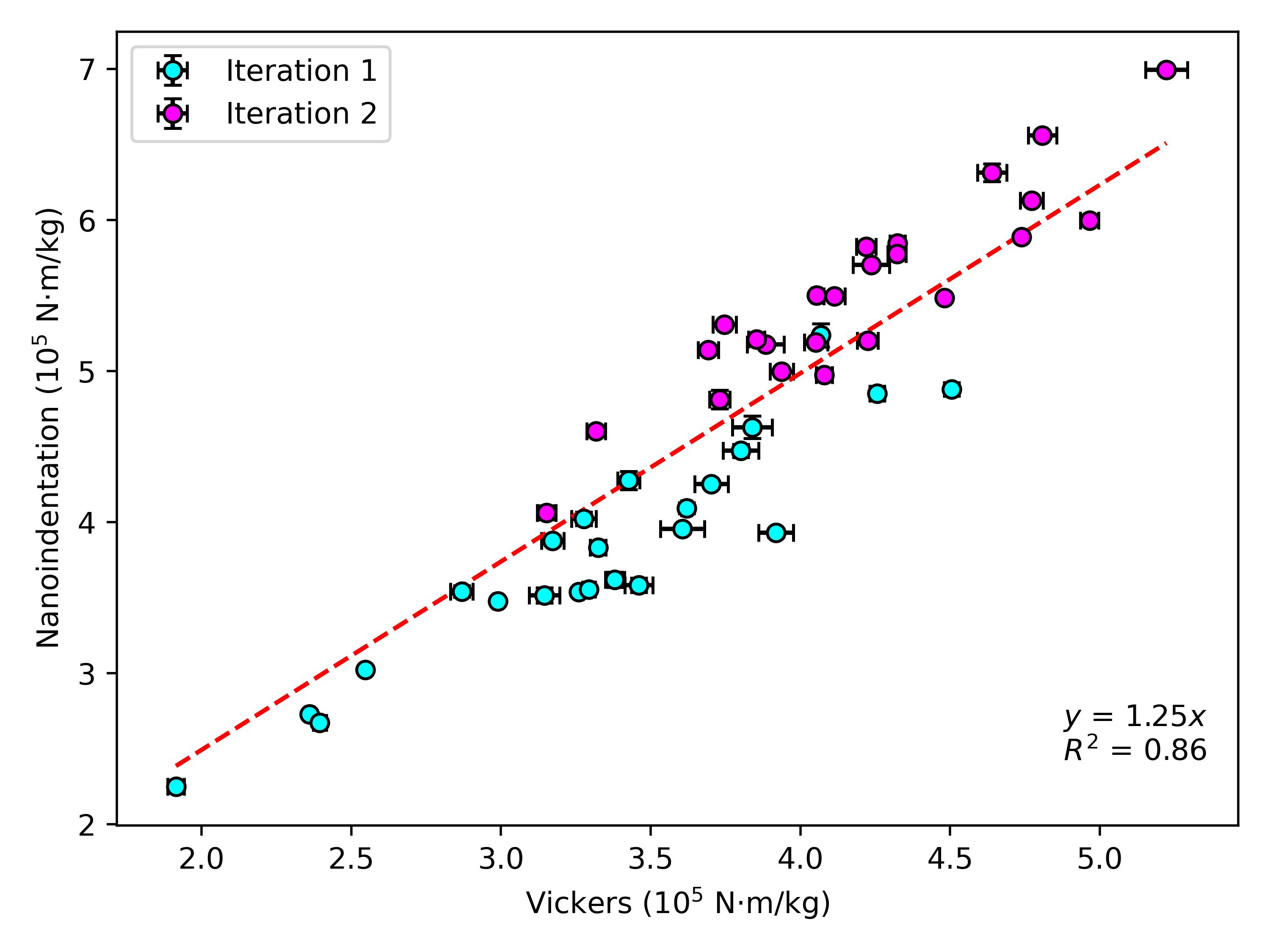}
    \caption{Comparison of specific hardness values measured via nanoindentation vs. Vickers microhardness for both iterations. Error bars represent one standard error. The dashed line, equation, and $R^2$ value correspond to a linear fit of the data.}
    \label{fig:NIvsVickers}
\end{figure}

As shown in Figure \ref{fig:NIvsVickers}, nanoindentation reported an approximately 25\% higher hardness than Vickers when averaged across all alloys in both iterations (i.e., a slope of 1.25 for a linear fit of data from the two methods). An easy answer to why this discrepancy exists is that nanoindentation is inherently performed on a smaller length scale and could, therefore, see enhanced hardness due to the indentation size effect \cite{Pharr2010}. However, a discerning reader may note that there are some minor differences in the rank ordering of hardnesses between the two methods. The full answer to why this discrepancy exists is complicated, as the two methods follow different standards and practices for performing indentation experiments and extracting hardness values \cite{Broitman2017}. For example, the Vickers method uses total contact area to measure hardness, whereas nanoindentation uses projected contact area. Everything else being equal, this would cause Vickers to report lower hardness values versus nanoindentation (i.e., contrary to the size effect). Vickers also results in a larger deformation volume within the material, leading to a potentially greater contribution to the measured hardness from surrounding grains. The comparison of these methods is further complicated by the fact that the Vickers method uses an indenter with four-fold symmetry, whereas nanoindentation typically uses a Berkovich indenter with three-fold symmetry. Therefore, the isotropy of the alloys and/or average orientation of the grains measured could affect the hardness values reported from one method versus the other.

Regardless of the differences in average values or even the specific rank ordering between select alloys, the general trend of hardness with respect to composition is consistent among the two methods. The utility of the BBO method used in this work is dependent on trends of the objective properties with respect to the input variables rather than the absolute objective property measurements. Therefore, these minor discrepancies are inconsequential to the optimization problem. The two indentation methods have their own advantages and limitations; thus, neither should be deemed inherently more reliable than the other within this context. Dissecting the subtle differences or exploring the nuanced implications of these variations falls outside the scope of this research. Nonetheless, the strong correlation of hardness trends with respect to alloy composition across both methods bolsters confidence in indentation as a reliable high-throughput characterization technique for evaluating the specific objective properties chosen in this work.

\subsection{Objective Property Optimization Results}
\label{Objective Property Optimization Results}

\begin{figure}
    \centering
    \includegraphics[width=1\linewidth]{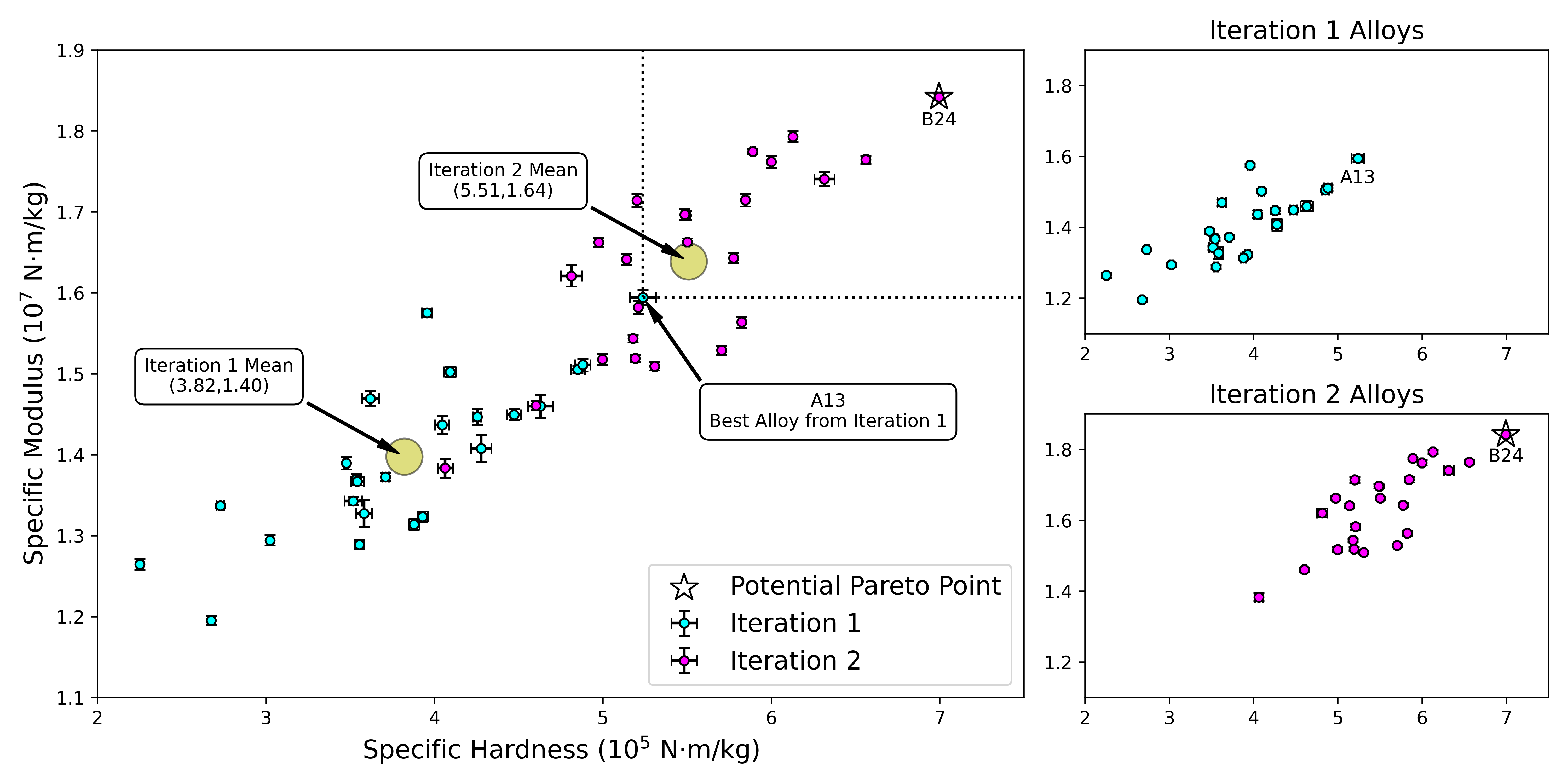}
    \caption{Objective property values and current potential Pareto point after both iterations (left). The plots on the right show the objective property data and best alloy from each iteration separately. Error bars represent one standard error.}
    \label{fig:Results}
\end{figure}

\begin{figure}
    \centering
    \includegraphics[width=0.5\linewidth]{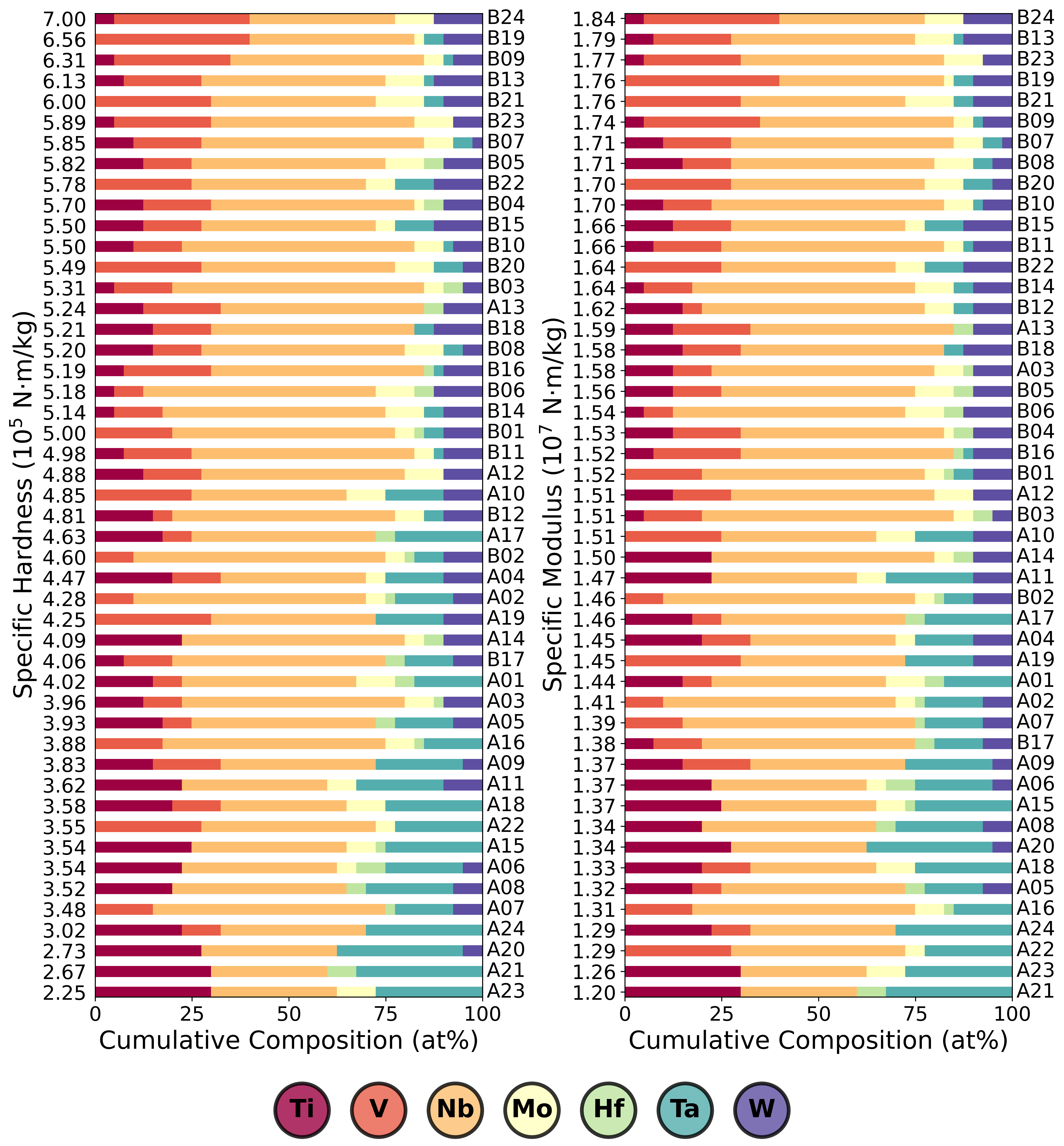}
    \caption{Rank ordering of all alloys produced by each objective property and their compositions. Colored bars correspond to the cumulative composition by each element, per the legend.}
    \label{fig:Rank_Order}
\end{figure}

The objective property results from both iterations are shown in Figure \ref{fig:Results}. The concept of the “Pareto front”, which is the set of “Pareto points” within an overall dataset, is useful when comparing data points within a multi-objective design problem. A Pareto point is a data point within a multidimensional space that is not dominated in \textit{all} dimensions by any other data point(s). When maximization is the goal, as done here, a Pareto point may have one dimension that is exceeded in value by other data points as long as none of those other data points \textit{also} exceed it in \textit{every other} dimension. The reader may note the authors’ use of the term “Potential Pareto Point” in Figure \ref{fig:Results}. This terminology was chosen to recognize the fact that identifying the \textit{true} Pareto front would require knowledge of the entire design space. Since only a small fraction of the design space has been queried after two iterations, the current best alloy (B24) only has the \textit{potential} to be a true Pareto point.

Among the first iteration alloys (Figure \ref{fig:Results}, top-right), we can see that only one alloy (A13) qualifies as a potential Pareto point. Notably, after two iterations, there was again only one alloy that qualifies as a potential Pareto point (B24). Though certainly not a rule, modulus and hardness often trend together among materials. Therefore, it is not entirely surprising that the alloy with the highest specific hardness also had the highest specific modulus after each iteration of synthesis and characterization. However, it should be realized that this was not an inevitability. If B24 was removed from the dataset, there would be two alloys qualifying as potential Pareto points (i.e., B19 and B21). Furthermore, if those two alloys were removed from the dataset, there would be a potential Pareto front with a population of three. As such, the fact that the potential Pareto front had a population of one after each iteration is a minor coincidence.

The selection of alloys for the first iteration was made using a presumption that a reasonably diverse set of inputs (i.e., compositions) would lead to significant optimizations in fewer iterations. One expects compositionally disparate materials to have disparate properties. Furthermore, providing the algorithm with a wide range of training data for how inputs relate to outputs should improve its efficiency in homing in on optimal regions of the design space. This was a primary motivation for using $k$-medoids clustering to select alloys widely distributed throughout the compositional space in each iteration, as discussed in sections \ref{Candidate Selection} and \ref{Batch Bayesian Optimization}. This strategy proved to be very effective, as discussed below.

As shown in Figure \ref{fig:Results} (top-right), mechanical characterization of the first iteration resulted in a $>$ 2-fold range in specific hardness and a $>$ 1.3-fold range in specific modulus. The wide range of both inputs (i.e., compositions) and objective properties in the first iteration translated to significant gains in the objective properties after only one round of optimization. The 24 alloys selected by the BBO algorithm for the second iteration far outperformed those selected in the first iteration. Specifically, 10 out of the 24 alloys produced in the second iteration dominated all of the alloys from the first iteration. Generally, one needs a hypervolume calculation to determine improvements in $n$-dimensional spaces \cite{Cao2015}. However, both iterations have a single dominant point. As such, one can reasonably evaluate improvement with merely a product of the objective properties, avoiding the expensive recursive function necessary to calculate hypervolume. If the objective property improvements are calculated using the product of the specific hardness and specific modulus of each alloy, the following property gains can be quantified. The sole potential Pareto point after the second iteration (B24) had a specific modulus $\times$ specific hardness product 54\% higher than the best alloy from the first iteration (A13). Furthermore, if the product of the mean specific hardness and mean specific modulus is assumed to be an indication of the overall performance of a particular iteration, the second iteration showed a 70\% improvement over the first iteration. 

Figure \ref{fig:Rank_Order} rank-orders all alloys among both iterations with respect to each objective property. Additionally, the relative atomic fractions of each element in each alloy are represented visually by the colored bars. A few notable qualities stand out from this visual: the alloys are largely niobium-rich, hafnium is mostly omitted, and better alloys tend to have more vanadium, tungsten, or niobium at the cost of titanium or tantalum. As a side note, it is a relatively happy coincidence that hafnium is sparse, being the most expensive of the alloying elements considered.

It is not surprising that hafnium is largely absent, as adding it in meaningful quantities quickly reduces the stability of a single-phase BCC microstructure. One could convince oneself of this fact by perusing binary phase diagrams between the alloying elements available in the literature \cite{ASM2016}. Furthermore, only molybdenum and niobium satisfy the density ($\rho < 11.39\ \mathrm{g/{cm}^3}$), melting point ($T_\mathrm{m}>2330$ °C), and crystal structure (BCC at 800 °C) constraints in elemental form. However, molybdenum was limited to $<$ 15 at\% for synthesis and processing considerations. Therefore, it is also not surprising that the alloys tended to contain large fractions of niobium. However, while these facts may seem relatively obvious in hindsight, relying on human intuition from the outset for a higher-dimensional optimization problem with multiple objective properties is inherently cumbersome and prone to bias. For example, the two alloys nearest to B24 in Figure \ref{fig:Results} (i.e., B19 and B21) do not contain titanium. The density of each alloy was used as both a constraint for filtering as well as part of the objective properties (i.e., each property was normalized by density). As such, one could be tempted to assume that the best alloys should all contain the alloying element with the lowest density (i.e., titanium). However, had that assumption been used to arbitrarily force the inclusion of titanium in all alloys, two of the three best alloys overall wouldn’t have been discovered. 

\section{Conclusions}

This work demonstrated rapid discovery and characterization within a compositionally complex refractory alloy space with explicit performance metrics (specific hardness and specific modulus). However, these techniques can be applied to any conceivable material space (metals, ceramics, polymers, composites, etc.) and to optimize any performance metric (mechanical, thermal, chemical, etc.) based on specific objectives outlined by the user. The feasibility of this framework for new materials discovery is only limited by the ability of the experimental methods employed to provide the optimization algorithm with reliable and quantifiable data for each of the design inputs and objective properties. 

The accomplishment of performing two full material design cycles in a few short months while demonstrating significant gains in the objective properties stands as a testament to this strategy in terms of effectively producing and testing new materials with the highest potential for success given the performance requirements. Furthermore, these significant gains were achieved with nothing about the objective properties vs. design inputs assumed \textit{a priori}. That is, the problem was approached with a completely blank slate; all decisions for initial sampling of the first iteration of alloys were made by the framework approaching a black box problem with respect to the objective properties. However, by using only the measured objective properties and compositional data from the first iteration, the optimization algorithm identified new alloys demonstrating substantial improvements in both objective properties. Specifically, the hypervolume of the best alloy produced in the second iteration was 54\% higher than that of the first iteration. Furthermore, 10 of the 24 alloys from the second iteration dominated all alloys from the first iteration with respect to both objective properties. Therefore, there was a clear indication that the algorithm was homing in on regions of the design space that optimized the objective properties with a surprisingly sparse amount of training data.

High-throughput computational thermodynamics and intelligent filtering were utilized in this work to downselect the alloy space using a range of prescribed property constraints. In any material discovery framework, pre-filtering the design space for feasible materials is an obvious step. For example, it would be foolish for the candidate space to include structural materials known or confidently predicted to melt below the operating temperature for which they are intended. Several constraints were chosen to demonstrate how such steps could be used to filter otherwise untenably large design spaces to more reasonable sizes for practical materials design problems. After such practicality and feasibility constraints were applied, an additional and purely statistical filtering step (i.e., $k$-medoids clustering) was used to reduce the candidate space to an experimentally tenable size of 1,000 candidate alloys. Finally, the first iteration of alloys was selected via an additional round of $k$-medoids clustering to select a sample that casted a wide net across the inputs, under the assumption that this would also produce a wide and representative sampling of the objective properties. This assumption was validated by the wide range of objective property values in the first iteration, which translated to substantial improvements in the second iteration.

The engine driving this materials discovery effort was the BIRDSHOT (Batch-wise Improvement in Reduced Design Space using a Holistic Optimization Technique) strategy, which takes advantage of batchwise experimental methods that are conducive to efficient processing and characterization workflows. Experiments are almost universally more economical with respect to both time and cost when batches of experiments can be conducted in parallel. Therefore, it is highly beneficial that a synergistic computational/experimental framework for rapid materials discovery is employed in a batchwise way. Furthermore, it is imperative that the experiments employed for generating data on the objective properties are both high-throughput and reliable. The primary impetus for employing a computational framework such as that discussed in this paper is the rapidity by which materials discovery can be achieved. The primary bottleneck for such a framework will often be the experimental work. Furthermore, if the experimental data is unreliable, the framework will be uprooted as the optimization algorithm will make correspondingly poor recommendations for future experiments. For these reasons, a substantial portion of this manuscript was dedicated to the experimental methods employed.

The underpinning of BIRDSHOT is the utilization of batch Bayesian optimization (BBO). Other data-driven optimization techniques, such as random forest and neural networks, are reliant on relatively large training datasets to produce meaningful predictions. However, obtaining large datasets of material properties via experimental means from a blank slate is time-consuming and difficult even with high-throughput experimental techniques. The beauty of BBO is the ability to make meaningful recommendations to improve the objective properties with minimal data. The underlying Gaussian process regressor and acquisition function used in BBO are more computationally expensive per data point evaluated than random forests or neural networks. However, when the candidate space has been efficiently filtered to a reasonably-sized feasible space and the training data are inherently limited due to their source in actual and unique experiments, the data-lean problem is particularly well-suited to the model-rich BBO technique. 

\section*{Acknowledgements}

This work was funded by the United States Army Futures Command University Technology Development Division (UTDD) via contract number W911NF-22-F-0032. Portions of this research were conducted with the advanced computing resources provided by Texas A\&M High Performance Research Computing. The authors also wish to acknowledge Danial Khatamsaz (Texas A\&M University) for extremely helpful guidance and discussion while developing the Bayesian optimization methods employed in this work. Some of the clipart representations included in Figure 1 were created with the assistance of artificial intelligence using OpenAI's ChatGPT-4.

\bibliographystyle{elsarticle-num} 
\bibliography{cas-refs}

\newpage

\section*{Supplemental Data}

\begin{table}[H]
    \centering
\caption{Nominal/target compositions, system type, and measured properties of all alloys produced. Alloys names beginning with "A" belong to the first iteration and alloy names beginning with "B" belong to the second iteration. Subsystem “size” and “designation” are discussed in section \ref{Initial Filtering}. Property ranges are given as calculated standard errors from experimental nanoindentation data.}
\label{tab:All Data}
\fontsize{8}{10}\selectfont
    \begin{tabular}{|l|c|c|c|c|c|c|c|c|c|c|c|} \hline 
         \textbf{Alloy}&\multicolumn{7}{|c|}{\textbf{Nominal Composition (at\%)}}&\multicolumn{2}{|c|}{\textbf{Subsystem}}&\textbf{Specific Hardness}&\textbf{Specific Modulus}
\\ \hline 
         &Ti&V&Nb&Mo&Hf&Ta&W&Size&Designation&$10^5$ N$\cdot$m/kg&$10^7$ N$\cdot$m/kg
\\ \hline 
        \textbf{A01}&  15.0&  7.5&  45.0&  10.0&  5.0&  17.5&  \cellcolor{grayshade}0.0&   6& 2&4.046 ± 0.042& 1.437 ± 0.011
\\ \hline 
        \textbf{A02}&  \cellcolor{grayshade}0.0&  10.0&  60.0&  5.0&  2.5&  15.0&  7.5&   6& 3&4.277 ± 0.061& 1.408 ± 0.017
\\ \hline 
        \textbf{A03}&  12.5&  10.0&  57.5&  7.5&  2.5&  \cellcolor{grayshade}0.0&  10.0&   6& 4&3.955 ± 0.030& 1.575 ± 0.005
\\ \hline 
        \textbf{A04}&  20.0&  12.5&  37.5&  5.0&  \cellcolor{grayshade}0.0&  15.0&  10.0&   6& 5&4.473 ± 0.042& 1.449 ± 0.007
\\ \hline 
        \textbf{A05}&  17.5&  7.5&  47.5&  \cellcolor{grayshade}0.0&  5.0&  15.0&  7.5&   6& 6&3.930 ± 0.030& 1.323 ± 0.006
\\ \hline 
        \textbf{A06}&  22.5&  \cellcolor{grayshade}0.0&  40.0&  5.0&  7.5&  20.0&  5.0&   6& 8&3.538 ± 0.024& 1.370 ± 0.006
\\ \hline 
        \textbf{A07}&  \cellcolor{grayshade}0.0&  15.0&  60.0&  \cellcolor{grayshade}0.0&  2.5&  15.0&  7.5&   5& 13&3.476 ± 0.012& 1.390 ± 0.008
\\ \hline 
        \textbf{A08}&  20.0&  \cellcolor{grayshade}0.0&  45.0&  \cellcolor{grayshade}0.0&  5.0&  22.5&  7.5&   5& 14&3.516 ± 0.051& 1.343 ± 0.005
\\ \hline 
        \textbf{A09}& 15.0& 17.5& 40.0& \cellcolor{grayshade}0.0& \cellcolor{grayshade}0.0& 22.5& 5.0& 5& 15& 3.709 ± 0.019&1.373 ± 0.005
\\ \hline 
        \textbf{A10}& \cellcolor{grayshade}0.0& 25.0& 40.0& 10.0& \cellcolor{grayshade}0.0& 15.0& 10.0& 5& 16& 4.851 ± 0.043&1.505 ± 0.004
\\ \hline 
        \textbf{A11}& 22.5& \cellcolor{grayshade}0.0& 37.5& 7.5& \cellcolor{grayshade}0.0& 22.5& 10.0& 5& 17& 3.620 ± 0.050&1.470 ± 0.009
\\ \hline 
        \textbf{A12}& 12.5& 15.0& 52.5& 10.0& \cellcolor{grayshade}0.0& \cellcolor{grayshade}0.0& 10.0& 5& 19& 4.881 ± 0.045&1.511 ± 0.008
\\ \hline 
        \textbf{A13}& 12.5& 20.0& 52.5& \cellcolor{grayshade}0.0& 5.0& \cellcolor{grayshade}0.0& 10.0& 5& 20& 5.237 ± 0.076&1.594 ± 0.009
\\ \hline 
        \textbf{A14}& 22.5& \cellcolor{grayshade}0.0& 57.5& 5.0& 5.0& \cellcolor{grayshade}0.0& 10.0& 5& 22& 4.093 ± 0.036&1.503 ± 0.006
\\ \hline 
        \textbf{A15}& 25.0& \cellcolor{grayshade}0.0& 40.0& 7.5& 2.5& 25.0& \cellcolor{grayshade}0.0& 5& 25& 3.542 ± 0.038&1.367 ± 0.005
\\ \hline 
        \textbf{A16}& \cellcolor{grayshade}0.0& 17.5& 57.5& 7.5& 2.5& 15.0& \cellcolor{grayshade}0.0& 5& 26& 3.878 ± 0.033&1.314 ± 0.007
\\ \hline 
        \textbf{A17}& 17.5& 7.5& 47.5& \cellcolor{grayshade}0.0& 5.0& 22.5& \cellcolor{grayshade}0.0& 5& 27& 4.629 ± 0.073&1.460 ± 0.014
\\ \hline 
        \textbf{A18}& 20.0& 12.5& 32.5& 10.0& \cellcolor{grayshade}0.0& 25.0& \cellcolor{grayshade}0.0& 5& 28& 3.583 ± 0.047&1.327 ± 0.017
\\ \hline 
        \textbf{A19}& \cellcolor{grayshade}0.0& 30.0& 42.5& \cellcolor{grayshade}0.0& \cellcolor{grayshade}0.0& 17.5& 10.0& 4& 31& 4.253 ± 0.023&1.447 ± 0.010
\\ \hline 
        \textbf{A20}& 27.5& \cellcolor{grayshade}0.0& 35.0& \cellcolor{grayshade}0.0& \cellcolor{grayshade}0.0& 32.5& 5.0& 4& 33& 2.729 ± 0.023&1.337 ± 0.003
\\ \hline 
        \textbf{A21}& 30.0& \cellcolor{grayshade}0.0& 30.0& \cellcolor{grayshade}0.0& 7.5& 32.5& \cellcolor{grayshade}0.0& 4& 41& 2.674 ± 0.008&1.195 ± 0.005
\\ \hline 
        \textbf{A22}& \cellcolor{grayshade}0.0& 27.5& 45.0& 5.0& \cellcolor{grayshade}0.0& 22.5& \cellcolor{grayshade}0.0& 4& 57& 3.554 ± 0.024&1.289 ± 0.006
\\ \hline 
        \textbf{A23}& 30.0& \cellcolor{grayshade}0.0& 32.5& 10.0& \cellcolor{grayshade}0.0& 27.5& \cellcolor{grayshade}0.0& 4& 58& 2.251 ± 0.021&1.265 ± 0.007
\\ \hline 
        \textbf{A24}& 22.5& 10.0& 37.5& \cellcolor{grayshade}0.0& \cellcolor{grayshade}0.0& 30.0& \cellcolor{grayshade}0.0& 4& 59& 3.024 ± 0.017&1.294 ± 0.006
\\ \hline 
        \textbf{B01}& \cellcolor{grayshade}0.0& 20.0& 57.5& 5.0& 2.5& 5.0& 10.0& 6& 3& 4.996 ± 0.011&1.518 ± 0.007
\\ \hline 
        \textbf{B02}& \cellcolor{grayshade}0.0& 10.0& 65.0& 5.0& 2.5& 7.5& 10.0& 6& 3& 4.603 ± 0.019&1.461 ± 0.005
\\ \hline 
        \textbf{B03}& 5.0& 15.0& 65.0& 5.0& 5.0& \cellcolor{grayshade}0.0& 5.0& 6& 4& 5.307 ± 0.010&1.509 ± 0.005
\\ \hline 
        \textbf{B04}& 12.5& 17.5& 52.5& 2.5& 5.0& \cellcolor{grayshade}0.0& 10.0& 6& 4& 5.704 ± 0.014&1.529 ± 0.006
\\ \hline 
        \textbf{B05}& 12.5& 12.5& 50.0& 10.0& 5.0& \cellcolor{grayshade}0.0& 10.0& 6& 4& 5.824 ± 0.013&1.564 ± 0.007
\\ \hline 
        \textbf{B06}& 5.0& 7.5& 60.0& 10.0& 5.0& \cellcolor{grayshade}0.0& 12.5& 6& 4& 5.178 ± 0.009&1.544 ± 0.005
\\ \hline 
        \textbf{B07}& 10.0& 17.5& 57.5& 7.5& \cellcolor{grayshade}0.0& 5.0& 2.5& 6& 5& 5.846 ± 0.011&1.715 ± 0.008
\\ \hline 
        \textbf{B08}& 15.0& 12.5& 52.5& 10.0& \cellcolor{grayshade}0.0& 5.0& 5.0& 6& 5& 5.201 ± 0.007&1.714 ± 0.008
\\ \hline 
        \textbf{B09}& 5.0& 30.0& 50.0& 5.0& \cellcolor{grayshade}0.0& 2.5& 7.5& 6& 5& 6.314 ± 0.060&1.741 ± 0.009
\\ \hline 
        \textbf{B10}& 10.0& 12.5& 60.0& 7.5& \cellcolor{grayshade}0.0& 2.5& 7.5& 6& 5& 5.495 ± 0.009&1.696 ± 0.005
\\ \hline 
        \textbf{B11}& 7.5& 17.5& 57.5& 5.0& \cellcolor{grayshade}0.0& 2.5& 10.0& 6& 5& 4.975 ± 0.006&1.663 ± 0.005
\\ \hline 
        \textbf{B12}& 15.0& 5.0& 57.5& 7.5& \cellcolor{grayshade}0.0& 5.0& 10.0& 6& 5& 4.813 ± 0.063&1.621 ± 0.013
\\ \hline 
        \textbf{B13}& 7.5& 20.0& 47.5& 10.0& \cellcolor{grayshade}0.0& 2.5& 12.5& 6& 5& 6.128 ± 0.010&1.793 ± 0.007
\\ \hline 
        \textbf{B14}& 5.0& 12.5& 57.5& 10.0& \cellcolor{grayshade}0.0& 5.0& 10.0& 6& 5& 5.139 ± 0.008&1.641 ± 0.007
\\ \hline 
        \textbf{B15}& 12.5& 15.0& 45.0& 5.0& \cellcolor{grayshade}0.0& 10.0& 12.5& 6& 5& 5.502 ± 0.010&1.663 ± 0.004
\\ \hline 
        \textbf{B16}& 7.5& 22.5& 55.0& \cellcolor{grayshade}0.0& 2.5& 2.5& 10.0& 6& 6& 5.192 ± 0.009&1.519 ± 0.005
\\ \hline 
        \textbf{B17}& 7.5& 12.5& 55.0& \cellcolor{grayshade}0.0& 5.0& 12.5& 7.5& 6& 6& 4.063 ± 0.046&1.384 ± 0.012
\\ \hline 
        \textbf{B18}& 15.0& 15.0& 52.5& \cellcolor{grayshade}0.0& \cellcolor{grayshade}0.0& 5.0& 12.5& 5& 15& 5.210 ± 0.011&1.582 ± 0.008
\\ \hline 
        \textbf{B19}& \cellcolor{grayshade}0.0& 40.0& 42.5& 2.5& \cellcolor{grayshade}0.0& 5.0& 10.0& 5& 16& 6.561 ± 0.010&1.764 ± 0.005
\\ \hline 
        \textbf{B20}& \cellcolor{grayshade}0.0& 27.5& 50.0& 10.0& \cellcolor{grayshade}0.0& 7.5& 5.0& 5& 16& 5.486 ± 0.008&1.697 ± 0.007
\\ \hline 
        \textbf{B21}& \cellcolor{grayshade}0.0& 30.0& 42.5& 12.5& \cellcolor{grayshade}0.0& 5.0& 10.0& 5& 16& 5.999 ± 0.008&1.762 ± 0.007
\\ \hline 
        \textbf{B22}& \cellcolor{grayshade}0.0& 25.0& 45.0& 7.5& \cellcolor{grayshade}0.0& 10.0& 12.5& 5& 16& 5.776 ± 0.013&1.643 ± 0.007
\\ \hline 
        \textbf{B23}& 5.0& 25.0& 52.5& 10.0& \cellcolor{grayshade}0.0& \cellcolor{grayshade}0.0& 7.5& 5& 19& 5.889 ± 0.014&1.775 ± 0.003
\\ \hline 
        \textbf{B24}& 5.0& 35.0& 37.5& 10.0& \cellcolor{grayshade}0.0& \cellcolor{grayshade}0.0& 12.5& 5& 19& 6.995 ± 0.012&1.842 ± 0.004
\\ \hline
    \end{tabular}

\end{table}

\begin{figure}
    \centering
    \includegraphics[width=1\linewidth]{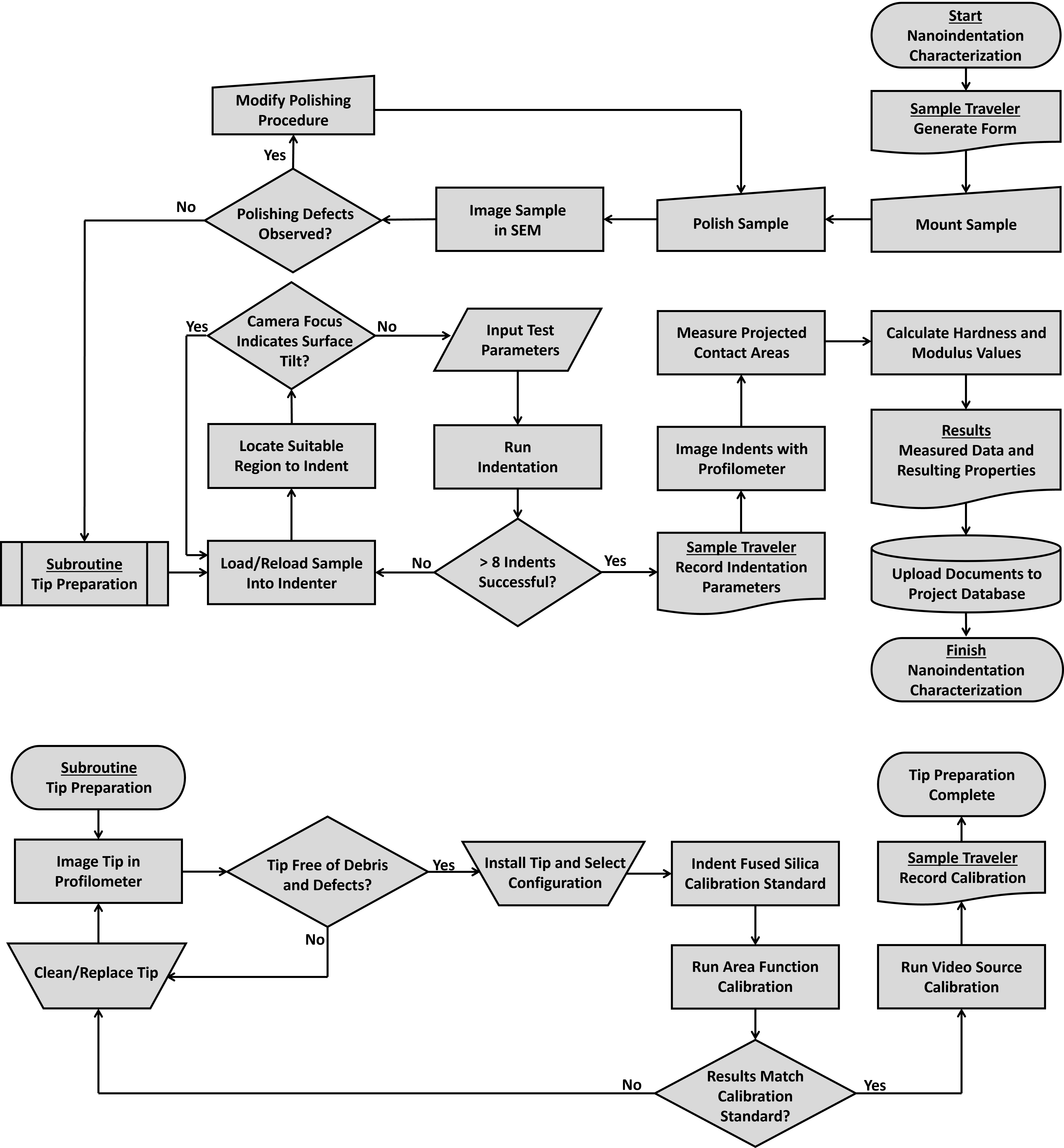}
    \caption{Decision tree used for the nanoindentation and indenter tip calibration procedures.}
    \label{fig:Decision_Tree}
\end{figure}

\begin{table}
    \centering
\caption{Indentation parameters used for all samples in both iterations.}
\label{tab:Nanoindentation Parameters}
    \begin{tabular}{ll}
    \hline
         Target Indent Depth (nm)& 2000
\\
         Maximum Load (N)& 1
\\
         Target Indentation Strain Rate ($\mathrm{s}^{-1}$)& 0.2
\\
         Target CSM Oscillation Frequency (Hz)& 110
\\
         Target CSM Dynamic Displacement (nm)& 2
\\
         Surface Approach Velocity (nm/s)& 150
\\
         Maximum Load Hold Time (s)& 1
\\
         Data Acquisition Rate (Hz)& 100
\\
    \hline
    \end{tabular}
\end{table}

\begin{figure}
    \centering
    \includegraphics[width=0.5\linewidth]{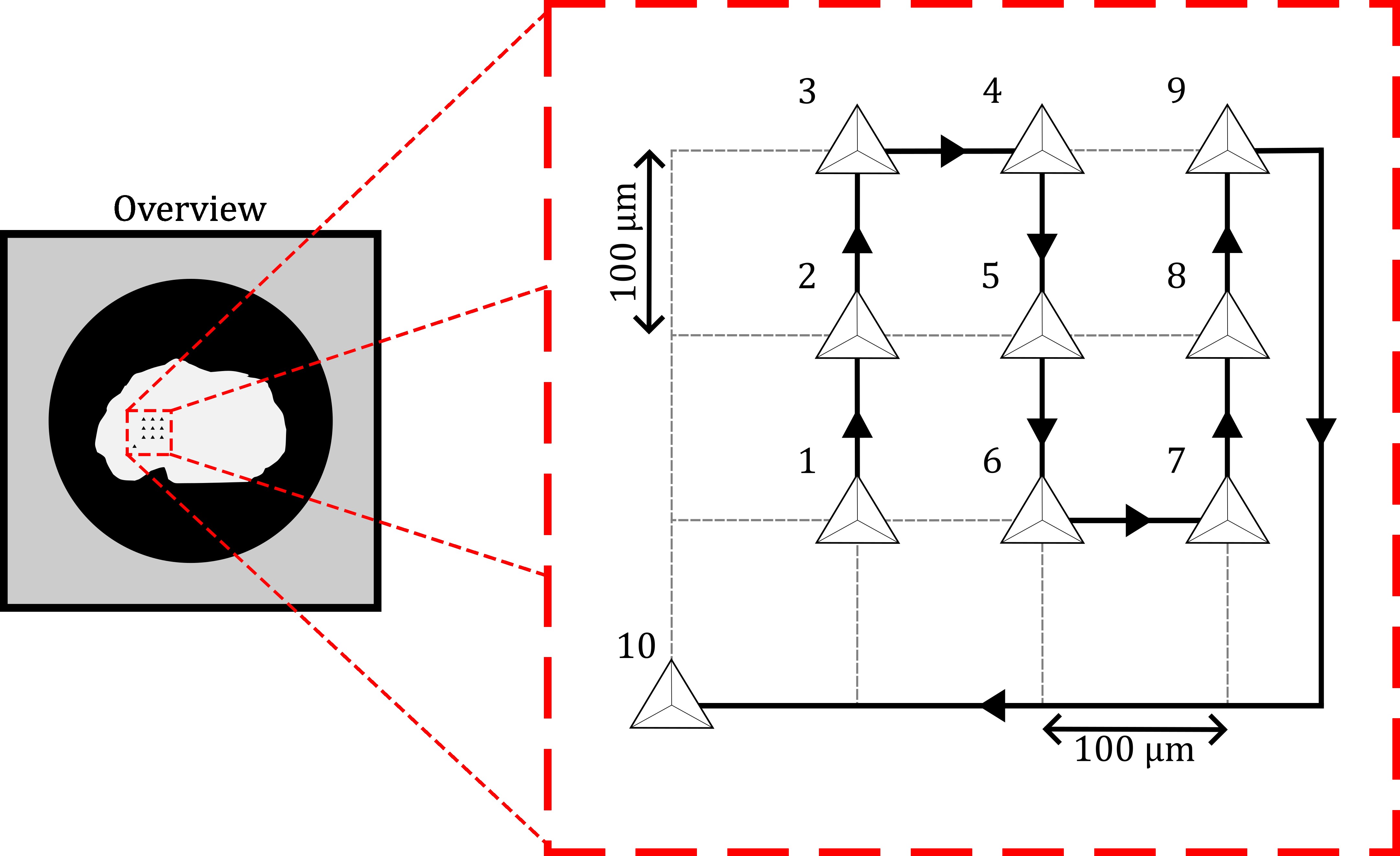}
    \caption{Schematic representation of the indent arrays used on each sample. Indent 10 acts as a fiducial marker to aid in post-indentation imaging for pile-up correction. Not to scale.}
    \label{fig:NI_Array}
\end{figure}

\end{document}